\documentclass[12pt,a4paper]{article}
\pdfoutput=1
 
\usepackage{jheppub-arxiv}
\usepackage[usenames,dvipsnames,table]{xcolor}
 
\usepackage{amssymb} 
\usepackage{amsmath}
\usepackage{mathtools}
\usepackage{amsfonts}    
\usepackage{dsfont}
\usepackage{pdfpages}
\usepackage{verbatim}
\usepackage{stmaryrd}
\usepackage{graphicx}
\usepackage{import}
\usepackage{etoc}
\usepackage{enumitem}
\usepackage{tensor}
\usepackage{mathrsfs}
\usepackage{textgreek} 
\usepackage[mathscr]{euscript}
\usepackage[smalltableaux]{ytableau}
\ytableausetup{centertableaux}
\usepackage{tikz}
\usetikzlibrary{decorations.pathreplacing, decorations.markings,calc,shapes.misc,decorations.pathmorphing,patterns.meta, math,positioning}
\usepackage{tabularx,ragged2e}
\usepackage{physics}

\usepackage{slashed}
\usepackage{datetime}
\usepackage{hyperref}
\hypersetup{
    pdfencoding=unicode,
	colorlinks=true,
	urlcolor=Maroon,
	linkcolor=RoyalBlue,
	citecolor=Maroon,
	pdftitle={Records from the S-Matrix Marathon: Observables in Expanding Universes},
	pdfauthor={},
	pdfdisplaydoctitle=true,
	pdfstartview=FitH,
	linktocpage=true
}

\usetikzlibrary{shapes}
\usepackage{enumitem}
\usepackage[export]{adjustbox}
\usepackage{nccmath}
\usepackage{bbm}
\usepackage{braket}
\usepackage{empheq}
\usepackage{cancel}
\usepackage{cleveref}
\usepackage[most]{tcolorbox}

\definecolor{charcoal}{HTML}{343837}

\usepackage{bm}
\definecolor{yellowish}{rgb}{0.880722,0.611041,0.142051}

\newcommand{\ba}{\begin{align}}

\newcommand{\be}{\begin{equation}}
\newcommand{\ee}{\end{equation}}
\def\bd{\begin{tikzpicture}}
\def\ed{\end{tikzpicture}}

\renewcommand\Re{\mathop{\text{Re}}}

\renewcommand{\vec}[1]{\mathbf{#1}}
\allowdisplaybreaks

\def\XXint#1#2#3{{\setbox0=\hbox{$#1{#2#3}{\int}$}
     \vcenter{\hbox{$#2#3$}}\kern-.5\wd0}}

\definecolor{light-gray}{gray}{0.75}

\newcommand\GL{\text{GL}}

\renewcommand\d{\text{d}}

\newcommand{\e}{\mathrm{e}}

\renewcommand{\L}{\mathrm{L}}

\renewcommand{\geq}{\geqslant}
\newcommand{\vx}{\vec{x}}
\newcommand{\vp}{\vec{p}}
\newcommand{\vell}{\bm{\ell}}
\newcommand{\rD}{\mathrm{D}}

\definecolor{dpurple}  {RGB} {189,  147,  249}

\usepackage{ntheorem}
\usepackage[framemethod=tikz,footnoteinside=false,hidealllines=true,topline=true,bottomline=true,linecolor=black!10!white,linewidth=1pt,innerleftmargin=0em,innerrightmargin=0em,frametitlerule=true,ntheorem]{mdframed}
\theorembodyfont{\upshape}

\newmdtheoremenv{mtheorem}{Theorem}[section]
\newmdtheoremenv[]{mdexample}{Example}[section]
\newmdtheoremenv{mdremark}{Remark}[section]
\newmdtheoremenv{mddefinition}{Definition}[section]
\newmdtheoremenv{mdcorollary}{Corollary}[section]
\newmdtheoremenv{mdproposition}{Proposition}[section]
\newmdtheoremenv{QA}{Audience question}[section]
\newcommand{\question}[1]{%
    #1\\[2ex] 
    \textit{Answer:} 
}

\title{{\Large\normalfont Records from the S-Matrix Marathon:}\\ Observables in Expanding Universes}

\author{{\normalfont Lecturers:}}
\author[1,2]{Paolo~Benincasa,}
\author[1]{Francisco~Vazão}

\author{\\ {\normalfont Notes written by:}}
\author[3]{Mathieu~Giroux,}
\author[4]{Holmfridur~S.~Hannesdottir,}
\author[4,5,6]{Sebastian~Mizera,}
\author[3]{Celina~Pasiecznik}

\affiliation[1]{Max-Planck-Institut f\"ur Physik, Werner-Heisenberg-Institut,\\ Boltzmannstrasse 8, D-85748 Garching, Germany}
\affiliation[2]{Instituto Galego de Física de Altas Enerxías IGFAE, Universidade de Santiago\\ de Compostela, E-15782 Galicia-Spain}
\affiliation[3]{Department of Physics, McGill University, 3600 Rue University,\\ Montr\'eal, H3A 2T8, QC Canada}
\affiliation[4]{Institute for Advanced Study, Princeton, NJ 08540, USA}
\affiliation[5]{Department of Physics, Princeton University, Princeton, NJ 08544, USA}
\affiliation[6]{Princeton Center for Theoretical Science,\\ Princeton University, Princeton, NJ 08544, USA}

\abstract{Observables in expanding universes are crucial to understand the physics of the early universe. In these lectures, we review some recent progress in understanding their mathematical structure and extract the physics encoded in them. After discussing the most salient features of an expanding background and their consequences for defining an observable, we focus on the so-called Bunch--Davies wavefunctional. We analyze its analytic properties on general grounds and introduce an integral representation for it in perturbation theory for a special class of scalar toy models.
We discuss both the diagrammatics associated to the usual Feynman rules as well as combinatorial rules on the graphs, which generate a representation free of spurious poles. Such combinatorial rules find their origin in the combinatorics of the {\it cosmological polytopes} of which we provide a gentle introduction to its definition and its main features. Finally, the combinatorics of the cosmological polytopes turns out to determine the combinatorics of a special class of {\it nestohedra} that encode the asymptotic behaviour of the cosmological integrals. We provide a general description of such structures and behaviour, which is of crucial importance to understand the infrared divergences which plague observables in an expanding background.

These notes are based on a series of lectures held during the S-Matrix Marathon workshop at the Institute for Advanced Study on 11--22 March 2024.
}

\begin{document}

\newtheorem{definition}{Definition}[section]
\newtheorem*{remark}{Remark}

\maketitle
\setcounter{page}{1}

\setcounter{tocdepth}{4}
\setcounter{secnumdepth}{4}

\makeatletter
\g@addto@macro\bfseries{\boldmath}
\makeatother

\newpage
\section*{Preface}

This article is a chapter from the \emph{Records from the S-Matrix Marathon}, a series of lecture notes covering selected topics on scattering amplitudes~\cite{RecordsBook}. They are based on lectures delivered during a workshop on 11--22 March 2024 at the Institute for Advanced Study in Princeton, NJ. We hope that they can serve as a pedagogical introduction to the topics surrounding the S-matrix theory.

These lecture notes were prepared by the above-mentioned note-writers in collaboration with the lecturers. 

\vfill
\section*{Acknowledgments}

P.B. would like to thank Dieter L{\"u}st and Gia Dvali for making his participation possible as well as the Galician Institute for High Energy Physics (IGFAE) and the University of Santiago de Compostela for hospitality. 
M.G.’s and C.P.'s work is supported in parts by the National Science and Engineering Council of Canada (NSERC) and the Canada Research
Chair program, reference number CRC-2022-00421. Additionally, C.P. is supported by the Walter C. Sumner Memorial Fellowship.
H.S.H. gratefully acknowledges funding provided by the J. Robert Oppenheimer Endowed Fund of the Institute for Advanced Study.
S.M.
gratefully acknowledges funding provided by the Sivian Fund and the Roger Dashen Member Fund at the Institute for Advanced Study. 
This material is based upon work supported by the U.S. Department of Energy, Office of Science, Office of High Energy Physics under Award Number DE-SC0009988.

The S-Matrix Marathon workshop was sponsored by the Institute for Advanced Study and the Carl P. Feinberg Program in Cross-Disciplinary Innovation.

\newpage

\section*{\label{ch:BenincasaVazao}Observables in Expanding Universes\\
\normalfont{\textit{Paolo Benincasa, Francisco Vazão}}}

\setcounter{section}{0}

\noindent\rule{\textwidth}{0.25pt}
\vspace{-0.8em}
\etocsettocstyle{\noindent\textbf{Contents}\vskip0pt}{}
\localtableofcontents
\vspace{0.5em}
\noindent\rule{\textwidth}{0.25pt}
\vspace{1em}

\noindent
Some references include those by D.~Anninos~\cite{Anninos:2012qw}, P.~Benincasa~\cite{Benincasa:2022gtd,benincasaLectures}, P.~Benincasa and F.~Vazão~\cite{Benincasa:2024lxe}.

\section[Generalities and perturbation theory]{Generalities and perturbation theory\\
\normalfont{\textit{Paolo Benincasa}}}
\label{sec:generalities_pert}

\subsection{Motivation}

The S-matrix has proven to be well suited to shed light on physics in various setups at appropriate energy scales, and the principles underlying it. In particle-collider experiments, for example, scattering amplitudes between ``in'' and ``out'' states provide a good approximation to experimental results. An essential assumption in the theoretical calculations is that the spacetime can be considered approximately flat for distances and times significantly greater than those characteristic of the scattering experiment. However, one of the most striking discoveries of the twentieth century is that our universe is currently undergoing a phase of accelerated expansion, so our spacetime is actually not asymptotically flat. In addition, the universe supposedly had a previous phase of \emph{inflation} -- an exponential expansion -- during its infancy. Inflation set the seeds for the subsequent evolution and encoded the cosmological structures we observe nowadays, such as temperature fluctuations in the cosmic microwave background. Said differently, the patterns that can be observed in the distribution of galaxies or in the temperature fluctuations in the cosmic microwave background can be traced back to observables at the end of inflation, which essentially provide the initial conditions for the post-inflationary era. Decoding such observables serves two purposes: understanding how the imprints of inflation manifest in structures today, and study the physics of inflation itself.

One reason for postulating an inflationary phase in the early universe is the need to address the \emph{horizon problem}, i.e.,  the question of why regions of space that are not causally connected are nevertheless observed to exhibit homogeneity. But yet  --- and despite its elegance in solving the horizon problem --- inflation remains to be better understood: the physics during inflation will not only unveil the phenomena that occurred in the very early universe, but more broadly provide a window into the physics at ``ultra high'' energies. Given that the Hubble parameter during inflation is estimated to be as large as $10^{14}$ GeV, it provides an opportunity to explore physics at energy scales that are much higher than those typical of colliders: 10-11 orders of magnitude greater than the energy that can be reached at the Large Hadron Collider at CERN and 5 orders of magnitude smaller than the Planck scale.

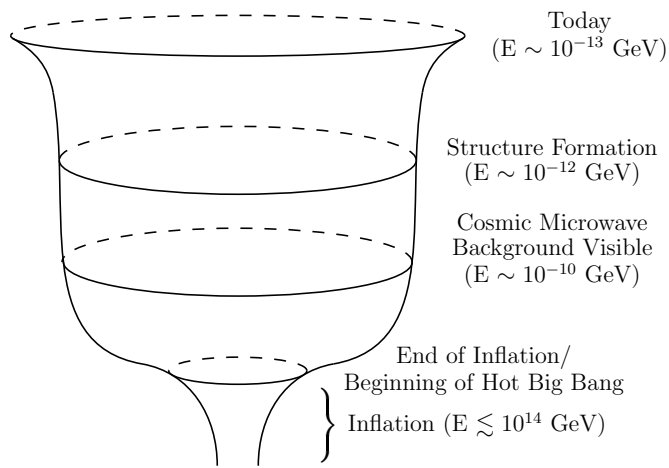
\begin{figure}[t]
    \centering
    \resizebox{0.6\textwidth}{!}{
   \begin{tikzpicture}[scale=1.2]
\draw (0.15,0.25) to[out=90,in=190] (0.7,1);
\draw (0.7,1) to[out=10,in=260] (1.35,3);
\draw (1.35,3) to[out=80,in=210] (1.648,3.4);
\draw[dashed] (0.505,0.95) arc[start angle=0,end angle=180,x radius=0.505,y radius=0.1];
\draw[] (0.505,0.95) arc[start angle=0,end angle=-180,x radius=0.505,y radius=0.1];
\node at (1.8,1.05) {\scalebox{0.5}{End of Inflation/}};
\node at (1.8,0.85) {\scalebox{0.5}{Beginning of Hot Big Bang}};
\draw[dashed] (1.275,1.75) arc[start angle=0,end angle=180,x radius=1.28,y radius=0.25];
\draw[] (1.275,1.75) arc[start angle=0,end angle=-180,x radius=1.275,y radius=0.25];
\node[align=center] at (2.3,2.05) {\scalebox{0.5}{Cosmic Microwave}};
\node[align=center] at (2.3,1.85) {\scalebox{0.5}{Background Visible}};
\node at (2.3,1.65) {\scalebox{0.5}{(E $\sim 10^{-10}$ GeV)}};
\draw[dashed] (1.305,2.5) arc[start angle=0,end angle=180,x radius=1.305,y radius=0.25];
\draw[] (1.305,2.5) arc[start angle=0,end angle=-180,x radius=1.305,y radius=0.25];
\node at (2.3,2.6) {\scalebox{0.5}{Structure Formation}};
\node at (2.3,2.4) {\scalebox{0.5}{(E $\sim 10^{-12}$ GeV)}};
\draw[dashed] (1.66,3.418) arc[start angle=0,end angle=180,x radius=1.66,y radius=0.15];
\draw[] (1.66,3.418) arc[start angle=0,end angle=-180,x radius=1.66,y radius=0.15];
\node at (2.5,3.518) {\scalebox{0.5}{Today}};
\node at (2.5,3.318) {\scalebox{0.5}{(E $\sim 10^{-13}$ GeV)}};
\draw (-0.15,0.25) to[out=90,in=350] (-0.7,1);
\draw (-0.7,1) to[out=170,in=280] (-1.35,3);
\draw (-1.35,3) to[out=100,in=330] (-1.648,3.4);
\node at (0.65,0.55) {\rotatebox{90}{\scalebox{0.5}{$\underbrace{\hspace{1.3cm}}$}}};
\node at (1.75,0.55) {\scalebox{0.5}{Inflation (E $\lesssim 10^{14}$ GeV)}};
\end{tikzpicture}}
    \caption{%
    Sketch of the evolution of our universe, where its snapshots relevant to our discussion are emphasised and the related energy scales. The infra-red physics of the large scale structure formation and the cosmic microwave background has its seeds in the initial conditions that result from the inflationary epoque.
    }
    \label{fig:Cosmo_energy_scales}
\end{figure}

Both the current and the inflationary expansions can be described in terms of an asymptotically quasi de Sitter (dS) spacetime, with a metric in the conformal global coordinates given by
\be
\d s^2_{1+d} = \left( \frac{\ell}{\cos T} \right)^2 [-\d T^2 + \d\theta^2 + \sin^2 \theta \, \d \Omega_{d-1}^2] \,,
\ee
where $T \in [-\pi/2, \pi/2]$, $\theta \in [0,\pi]$ and $d$ is the number of the spatial dimensions.\footnote{Here, we denote the number of spatial dimensions with $d$, which is different from the total number of space-time dimensions, $\rD=d+1$, used in other chapters.}
In order to understand the properties of this metric it is useful to draw its Penrose diagram that depicts an observer in an expanding universe surrounded by a cosmic horizon,
\begin{equation}
\adjustbox{valign=c}{
\begin{tikzpicture}
    \node[white] at (4.9,0){.};
    \coordinate (a) at (0,0);
    \coordinate (b) at (0,2);
    \coordinate (c) at (2,0);
    \coordinate (d) at (2,2);
    \coordinate (e) at (1,1);
    \fill[gray!70] (a) -- (b) -- (e) -- cycle;
    \draw (a) to (b);
    \draw (a) to node[below, pos=0.5] {$\mathscr{I}^-$} (c);
    \draw (b) to node[above, pos=0.5] {$\mathscr{I}^+$} (d);
    \draw (c) to (d);
    \draw (a) to (d);
    \draw (b) to (c);

    \draw[Maroon,ultra thick] (a) to (e);
    \draw[thick, ->,  >=stealth,bend left,Maroon] (0.5,0.5) to (-0.1,-1.2);
    \node[left, text=Maroon] at (-0.1,-1.2) {Cosmic horizon};
    \draw[thick, ->,  >=stealth,bend left,black] (0.6,0.9) to (-0.1,-1.6);
    \node[left] at (-0.1,-1.6){Causal access};

    \node[right] at ($(d) + (0,-0.2)$) {\scalebox{0.5}{N}};
    \node[right] at ($(d) + (0,-0.4)$) {\scalebox{0.5}{O}};
    \node[right] at ($(d) + (0,-0.6)$) {\scalebox{0.5}{R}};
    \node[right] at ($(d) + (0,-0.8)$) {\scalebox{0.5}{T}};
    \node[right] at ($(d) + (0,-1.0)$) {\scalebox{0.5}{H}};
    \node[right] at ($(d) + (0,-1.2)$) {\scalebox{0.5}{P}};
    \node[right] at ($(d) + (0,-1.4)$) {\scalebox{0.5}{O}};
    \node[right] at ($(d) + (0,-1.6)$) {\scalebox{0.5}{L}};
    \node[right] at ($(d) + (0,-1.8)$) {\scalebox{0.5}{E}};
    
    \node[left] at ($(b) + (0,-0.2)$) {\scalebox{0.5}{S}};
    \node[left] at ($(b) + (0,-0.4)$) {\scalebox{0.5}{O}};
    \node[left] at ($(b) + (0,-0.6)$) {\scalebox{0.5}{U}};
    \node[left] at ($(b) + (0,-0.8)$) {\scalebox{0.5}{T}};
    \node[left] at ($(b) + (0,-1.0)$) {\scalebox{0.5}{H}};
    \node[left] at ($(b) + (0,-1.2)$) {\scalebox{0.5}{P}};
    \node[left] at ($(b) + (0,-1.4)$) {\scalebox{0.5}{O}};
    \node[left] at ($(b) + (0,-1.6)$) {\scalebox{0.5}{L}};
    \node[left] at ($(b) + (0,-1.8)$) {\scalebox{0.5}{E}};

    \node [left, xshift=-6] at (b) {$\frac{\pi}{2}$};
    \node [left, xshift=-6] at (a) {$-\frac{\pi}{2}$};
    \node [below,yshift=-3] at (a) {$\pi$};
    \node [below,yshift=-3] at (c) {$0$};

    \node[left] at (-0.7,1) {Observer};
    \draw[->,  >=stealth,thick] (-0.6,0.7) -- (-0.6,1.3);
    \node at (1,-1) {$\theta$};
    \draw[->,  >=stealth,thick] (1.2,-0.6) -- (0.8,-0.6);
\end{tikzpicture}
}
\end{equation}
In this diagram, each point is a $(d{-}1)$-dimensional sphere at a fixed value of the coordinate $\theta$, except at the points $\theta=0$ and $\theta=\pi$ (referred to as the north and south pole, respectively), where the sphere shrinks to a point. A light signal travels at 45$^\circ$ which means that an observer at the south pole, for example, can receive signals from another observer at $\mathscr{I}^-$ (the infinite past) and can send signals to a future observer at $\mathscr{I}^+$ (the infinite future). But that implies that the south-pole observer only has full causal access to the shaded southern causal diamond, i.e.,  they can send and receive signals only to points within the shaded region in the diagram.
 
An observer in an expanding universe is thus surrounded by a \emph{cosmic horizon}: the boundary of the region of spacetime to which they have full causal access. Importantly, the cosmic horizon is observer dependent, so there is not a universal notion of a boundary where our observer can be placed to make observations. In general, the data outside the cosmic horizon do not have a clear physical meaning. In a theory of gravity it is not possible to define local observables, as concentrating too much energy in a finite-sized region of spacetime can cause gravitational backreaction and eventually the formation of black holes, obstructing the ability to perform arbitrarily precise measurements. One can be tempted to consider data at the boundary of the spacetime, as we do in flat space, where we can define an S-matrix. However, this is not possible in a cosmological space-time as, in this case, such data lie outside the observer's horizon. 

What partially saves us is that inflation ends at a certain time. Let us consider a constant time-slice in the Poincar\'e patch, which covers half of dS and is the patch relevant for cosmology. Its metric is
\begin{align}
\d s^2 & = - \d t^2 + \e^{2t/\ell}[r^2 \d \theta^2 + \sin^2 \theta \d \Omega_{d-1}^2]
\nonumber
\\
& = (-\ell/\eta)^2 [-\d\eta^2 + \d\vec{x}^2] \,,
\end{align}
where $ t \in \mathbb{R}$ is the physical time and $\eta \in \mathbb{R}_-$ is the conformal time. This patch corresponds to the upper triangle in the de Sitter Penrose diagram, 
\begin{equation}
\adjustbox{valign=c}{
    \begin{tikzpicture}[scale=1.2]
    \coordinate (a) at (0,0);
    \coordinate (b) at (0,2);
    \coordinate (c) at (2,0);
    \coordinate (d) at (2,2);
    \coordinate (e) at (0,1.2);
    \draw (a) to (b) ;
    \draw (a) to node[below, pos=0.5] {$\mathscr{I}^-$} (c);
    \draw (b) to node[above, pos=0.5] {$\mathscr{I}^+$} (d);
    \draw[] (c) to (d);
    \draw[dashed, thick] (a) -- (d) node[midway, sloped, below] {$\eta \to -\infty$};
    \draw[dashed,thick] (b) to (c);
    \draw[thick] (d) [out=220,in=10]  to (e);
    \node[left] at (e) {$\eta_0$};
\end{tikzpicture}}
\end{equation}
At a given time, $\eta = \eta_0$, an observer can have access to a region determined by $\eta_0$, while if they wait long enough, i.e.,  take $\eta_0 \to 0^-$, all that data becomes data at $\mathscr{I}^+$. This implies that an observer has access to a finite amount of data, and the precision of the measurements are constrained by the size of the data set. Hence, the measurements carried out by a de Sitter observer at any finite time have a classical uncertainty, 
and, as a consequence, there is a difficulty in measuring a classical quantity even in principle. The key point is:
\begin{equation}
    \boxed{\text{Any measurement has a classical uncertainty.}}
\end{equation}
In other words, one cannot make arbitrarily precise measurements in a de Sitter spacetime.

\begin{QA}
\question{How does the error on classical uncertainty scale?}
One can, e.g., measure a massive scalar two-point function $\langle \phi \phi \rangle \sim \eta_0^\Delta$. Then, the uncertainty on $\Delta$ can be quantified as roughly $\sqrt{N}$ where $N$ is the number of modes one has access to. For example, if one takes a wavelength of the size of the universe, there will be a very large uncertainty since we can only make one measurement, but with smaller wavelengths one can make more measurements. 
\end{QA}

However, assuming that the universe at late enough time becomes infinitely large and flat, then it is possible to define equal-time correlation functions on such spacelike (late time) surfaces,
\begin{equation}
\adjustbox{valign=c}{
\begin{tikzpicture}[scale=1.4]
\draw (0.15,0.25) to[out=90,in=190] (0.7,1);
\draw (0.7,1) to[out=10,in=260] (1.35,3);
\draw (1.35,3) to[out=80,in=210] (1.648,3.4);
\draw[dashed] (0.505,0.95) arc[start angle=0,end angle=180,x radius=0.505,y radius=0.1];
\draw[] (0.505,0.95) arc[start angle=0,end angle=-180,x radius=0.505,y radius=0.1];
\draw[dashed] (1.275,1.75) arc[start angle=0,end angle=180,x radius=1.28,y radius=0.25];
\draw[] (1.275,1.75) arc[start angle=0,end angle=-180,x radius=1.275,y radius=0.25];
\draw[dashed] (1.305,2.5) arc[start angle=0,end angle=180,x radius=1.305,y radius=0.25];
\draw[] (1.305,2.5) arc[start angle=0,end angle=-180,x radius=1.305,y radius=0.25];
\draw[dashed] (1.66,3.418) arc[start angle=0,end angle=180,x radius=1.66,y radius=0.15];
\draw[] (1.66,3.418) arc[start angle=0,end angle=-180,x radius=1.66,y radius=0.15];
\draw (-0.15,0.25) to[out=90,in=350] (-0.7,1);
\draw (-0.7,1) to[out=170,in=280] (-1.35,3);
\draw (-1.35,3) to[out=100,in=330] (-1.648,3.4);

\draw[decoration={markings, mark=at position 0.5 with {\arrow[scale=0.8]{stealth}}}, postaction={decorate}] (0,0.89) to (0.3,0.95);
\draw[decoration={markings, mark=at position 0.7 with {\arrow[scale=0.8]{stealth}}}, postaction={decorate}] (0.3,0.95) to (0,1);
\draw[decoration={markings, mark=at position 0.5 with {\arrow[scale=0.8]{stealth}}}, postaction={decorate}] (0,1) to (-0.3,0.95);
\draw[decoration={markings, mark=at position 0.7 with {\arrow[scale=0.8]{stealth}}}, postaction={decorate}] (-0.3,0.95) -- (0,0.89);

\draw[decoration={markings, mark=at position 0.55 with {\arrow{stealth}}}, postaction={decorate}] (0,1.55) to (0.6,1.75);
\draw[decoration={markings, mark=at position 0.6 with {\arrow{stealth}}}, postaction={decorate}] (0.6,1.75) to (0,1.95);
\draw[decoration={markings, mark=at position 0.55 with {\arrow{stealth}}}, postaction={decorate}] (0,1.95) to (-0.6,1.75);
\draw[decoration={markings, mark=at position 0.55 with {\arrow{stealth}}}, postaction={decorate}] (-0.6,1.75) to (0,1.55);

\draw[decoration={markings, mark=at position 0.55 with {\arrow{stealth}}}, postaction={decorate}] (0,2.3) to (0.6,2.5);
\draw[decoration={markings, mark=at position 0.6 with {\arrow{stealth}}}, postaction={decorate}] (0.6,2.5) to (0,2.7);
\draw[decoration={markings, mark=at position 0.55 with {\arrow{stealth}}}, postaction={decorate}] (0,2.7) to (-0.6,2.5);
\draw[decoration={markings, mark=at position 0.55 with {\arrow{stealth}}}, postaction={decorate}] (-0.6,2.5) to (0,2.3);

\node[right] at (1.3,1) {$\langle\Phi \Phi \ldots\rangle$};
\node[right] at (1.3,1.75) {$\langle\delta T \delta T \ldots\rangle$};
\node[right] at (1.3,2.5) {$
\left\langle\delta \rho_g \delta \rho_g \ldots\right\rangle$};

\node[left] at (-1.35,1) {$t \sim 10^{-32}$ s};
\node[left] at (-1.35,1.75) {$t \sim 10^5$ yrs};
\node[left] at (-1.35,2.5) {$t \sim 10^{10}$ yrs};
\end{tikzpicture}}
\end{equation}
That is, we use this approximation for certain types of modes for which the late slice can be approximated as being at future infinity.
One can look for patterns in, for example, the density distribution of galaxies in the temperature fluctuations in the cosmic microwave background. These correlations are the result of the evolution from the initial conditions, which are set by inflation. In other words, the inflationary evolution process encoded in the correlations $\langle \Phi \Phi \cdots \rangle$ at the end of inflation serves as initial conditions for the temperature fluctuation correlations $\langle \delta T \delta T \cdots \rangle$ in the cosmic microwave background and the galaxy distributions, $\langle \delta \rho_g \delta \rho_g \cdots \rangle$. The focus of the following discussion will be on the correlations at the end of inflation $\langle \Phi \Phi \cdots \rangle$ since  
\begin{itemize}
    \item[(i)] as already mentioned, they are the initial conditions that led to the actual observables,
    \item[(ii)] they are the result of the inflationary evolution and thus they should encode inflationary physics,
    \item[(iii)]
    their full-fledged determination might allow us to make precise a notion of of holography and, thus, have a formulation of the microscopic theory.
\end{itemize}

\subsection{Observables: Correlators, probability distributions, wavefunctions}

A general definition of an equal-time correlation function is given as an integral over a certain probability distribution,
\be
\langle \Phi(\vec{p}_1) \cdots \Phi(\vec{p}_n) \rangle = \int \mathcal{D} \Phi P[\Phi] \Phi(\vec{p}_1) \cdots \Phi(\vec{p}_n),
\label{eq:phi_exp}
\ee
where $P[\Phi]$ is a probability distribution
satisfying $\int \mathcal{D} \Phi P[\Phi] =1$. We choose to consider the operator $\Phi(\vec{p}_1) \cdots \Phi(\vec{p}_n)$, but~\eqref{eq:phi_exp} holds for any quantity $f[\Phi]$ modulo the replacement $\Phi(\vec{p}_1) \cdots \Phi(\vec{p}_n) \to f[\Phi]$ on both sides.

If we assume that the full system can be described via a wavefunctional $\Psi[\Phi]$, then
\be
P[\Phi] = \frac{|\Psi[\Phi]|^2}{\displaystyle\int \mathcal{D} \Phi |\Psi[\Phi]|^2}\,,
\ee
where the wavefunctional is nothing but the transition amplitudes from a vacuum state at past infinity and a state $\Phi$ at the quasi-dS space-like future boundary located at $\eta\longrightarrow0^{-}$:
\be
\Psi[\Phi] = \langle \Phi | \hat{U}(0,-\infty) | 0 \rangle,
\ee 
with the evolution operator $\hat{U}(0, -\infty)$ defined as
\be
\hat{U}(0, -\infty) = \hat{T} \exp \{ - i \int_{-\infty}^{0} \d \eta \, H(\eta)\} \,.
\ee
As the correlations are computed on a fixed time-slice -- either future infinity at $\eta=0$ or, in case inflation ends before getting to the future boundary, at some small but finite $\eta=\eta_{\circ}$ --, they can pictorially be represented as: 
\begin{equation}
\adjustbox{valign=c}{
    \begin{tikzpicture}
    \coordinate (a) at (0,0);
    \coordinate (b) at (1,0);
    \coordinate (c) at (1.5,0);
    \coordinate (e) at (3.2,0);
    \coordinate (d) at (4,0);
    \coordinate (h) at (4,-2);
    \coordinate (f) at (2.7,0);
    \coordinate (g) at (2,0);
    \coordinate (ar1) at (0,-2);
    \coordinate (ar2) at (0,-0.1);
    \coordinate (cen) at (2,-1);
    \draw[very thick] (a) -- (d) node[right] {$\eta=0$};
    \draw[thick,fill=black] (b) circle (1pt) (c) circle (1pt) (e) circle (1pt) (f) circle (1pt) (g) circle (1pt);
    \draw[thick] (cen) to [out=180,in=-90] (b);
    \draw[thick] (cen) to [out=180,in=-90] (c) ;
    \draw[thick] (cen) to [out=0,in=-90] (e);
    \draw[thick] (cen) to [out=0,in=-90] (f);
    \draw[thick] (cen) to [out=90,in=-90] (g);
    \draw[black!30,fill=black!30] (cen) circle (0.5);
    \draw[->, thick] (ar1) node[right,yshift=-10] {$\eta \to -\infty$} -- (ar2);
\end{tikzpicture}}
\end{equation}
So far, the definition is pretty general, with the vacuum state yet to be specified. Importantly, the evolution operator is taken to be unitary (and hence $H(\eta)$ to be Hermitian).

Note that expressing the probability distribution in terms of the wavefunctional $\Psi[\Phi]$, the correlator acquires the usual Schwinger--Keldysh form,
\begin{equation}
    \langle \Phi \cdots \Phi \rangle = \mathcal{N} \int \mathcal{D} \Phi \, \langle 0 | [U(0,-\infty)]^\dagger | \Phi \rangle \Phi \cdots \Phi \langle \Phi | U(0,-\infty) | 0 \rangle \,.
\end{equation}
Interestingly, such a formula hints to the fact that the correlation function must inherit some properties of the wavefunctional as well as (at least) a subset of its properties derived from the properties of the wavefunctional. As any average can be computed from a probability distribution expressed in terms of the wavefunctional, the wavefunctional can be considered as a more primitive object. It will be the focus of most of our discussion.

It can be expressed in the more useful form of a path integral,
\be
\Psi[\Phi] = \mathcal{N} \int_{\phi(-\infty)=0}^{\phi(0) = \Phi} \mathcal{D} \phi \, \e^{iS[\phi]}\,,
\ee
where $S[\phi]$ is an action expressed in therms of the collection of the bulk fields $\phi(\eta,\vx)$, with boundary conditions taken to be the vacuum in the infinite past, and $\Phi(\vx)$ at $\mathscr{I}^+$.

The usual choice for a vacuum is the \emph{Bunch--Davies vacuum}, i.e.,  the one that is exponentially suppressed as $\eta \to -\infty$ while selecting positive-frequency solutions only,
\be
\phi(\eta, \vec{x}) \xrightarrow[\eta \to -\infty]{} f(\eta) \e^{i E \eta}\,,
\label{eq:phi_BD}
\ee
with $E = | \vp |$ and $f$ is a function which depends on the background (i.e.,  encodes the cosmology). The factor $\e^{ i E \eta}$ shows positive frequencies only and decays exponentially upon suitable regularization; e.g.,\ $\eta \to \eta (1- i \varepsilon)$ or $E \to E - i\varepsilon$, with $\varepsilon>0$.

\subsection{A comment on the \texorpdfstring{$i \varepsilon$}{} prescription}

The usual prescription for regulating the wavefunctional as early times are approached is to deform the integration contour around $-\infty$,
\begin{equation}
        \hat{U} \to \hat{U}_\varepsilon = \hat{T} \exp \left\{ - i \int_{-\infty (1-i \varepsilon)}^0 \d \eta \, H(\eta) \right\}\,.
\end{equation}
However, notice that such a deformation breaks unitarity as $\hat{U}_\varepsilon^\dag \neq \hat{U}_\varepsilon^{-1}$. A way to obtain convergence while preserving unitarity is via deforming the Hamiltonian rather than the integration contour, while preserving its Hermiticity \cite{Baumgart:2020oby}:
\begin{equation}
    H(\eta) \to\e^{\varepsilon \eta} H(\eta) \,.
\end{equation}
More generally, one can give a small negative imaginary part to the energies. One can take different a $\varepsilon$ for each energy, $E_j \to E_j - i\varepsilon_j$, $\varepsilon_j>0$ \cite{Albayrak:2023hie}. The above deformation of the Hamiltonian is just a special case of such a deformation.

\begin{QA}
    \question{What happens in the limit as $\varepsilon \to 0$?}
    Unitarity is restored in the strict limit $\varepsilon \to 0$. Nevertheless, the prescription above is important for manifestly recovering the flat-space limit of the cosmological cutting rules, as well as non-perturbatively.
\end{QA}

\subsection{Perturbative rules for the Bunch--Davies wavefunction}
The path integral formulation of the wavefunctional allows us to easily derive the Feynman rules in perturbation theory. In particular, if we split $\phi$ into a classical free part $\phi_{\text{cl}}$ and its fluctuations $\varphi$,
\begin{equation}
    \phi \to \phi_{\text{cl}} + \varphi \,,
\end{equation}
then, 
\be
\psi[\Phi] = \mathcal{N} \int_{\phi(-\infty)=0}^{\phi(0) = \Phi} \mathcal{D} \phi \,  \e^{i S[\phi]} = \e^{i S_{2}[\phi_{\text{cl}}]} \mathcal{N}'  \int_{\varphi(-\infty)=0}^{\varphi(0)=0} \mathcal{D} \varphi \, \e^{i S_2[\varphi]} \e^{i S_{\text{int}}[\phi_{\text{cl}}, \varphi]}\,,
\ee
where $S_2$ indicates the free action  -- the part that is evaluated at the classical solution while the integral is just over the fluctuations. The factors $\mathcal{N}$ and $\mathcal{N}'$ are normalizations. Importantly, the boundary condition of the fluctuations made them vanish both at the infinite past and at $\mathscr{I}^+$. This is easy to understand since 
\begin{align}
    \phi(\eta=0) = \Phi  \quad & \Rightarrow \quad \phi_{\text{cl}} (\eta=0) =  \Phi\,, \qquad \varphi (\eta=0) = 0\,, \\
    \phi(\eta\to -\infty) = 0 \quad & \Rightarrow \quad \phi_{\text{cl}} (\eta\to -\infty ) =  0\,, \qquad \varphi (\eta \to -\infty ) = 0\,,
\end{align}
expanding in powers of $S_{\text{int}}$, the path-integral expression can be organized as
\begin{equation}
    \begin{split}
            \Psi[\Phi] & = \exp \Bigg\{- \prod_{j=1}^{2} \left[ \int \frac{\d^d \vp_j}{(2\pi)^d } \Phi(\vp_j) \right] \psi_2(\vp_1, \vp_2) \Bigg\}
    \\ & \hspace{2cm} \times
    \left\{ 1 + \sum_{n > 0} \int  \prod_{k=1}^{n} \left[ \frac{\d^d \vp_k}{(2\pi)^d} \Phi(\vp_k) \right] \psi_n(\vp_1, ..., \vp_n) \right\}\,,
    \end{split}
\end{equation}
with the ``wavefunction coefficients'' $\psi_n(\vp_1,...,\vp_n)$, which can be expressed in terms of Feynman graphs:
\begin{itemize}
    \item[(i)] $\psi_2$ appearing in the factorized exponential is just the free two-point wavefunction and arises from the boundary term of the free on-shell action. Notice that the Feynman graphs are characterized by having their external states ending on a horizontal line that represents the future spacelike boundary $\mathscr{J}^+$ at $\eta=0$; the two-point wavefunction, for example, has support on $\delta^{(d)}(\vp_1+\vp_2)$.
    \be
        \begin{gathered}
    \begin{tikzpicture}[baseline={([yshift=2ex]current bounding box.center)},scale=1.0001]
    \coordinate (a) at (0,0);
    \coordinate (b) at (1,0);
    \coordinate (c) at (2,0);
    \coordinate (d) at (4,0);
    \coordinate (ar1) at (-0.5,-2);
    \coordinate (ar2) at (-0.5,-0.5);
    \draw[very thick] (a) -- (d) node[right] {$\eta=0$};
    \draw[thick,fill=black] (b) circle (1pt) (c) circle (1pt);
    \draw[very thick] (b) to [in=-90,out=-90, looseness=2] (c);
    \draw[->, thick] (ar1) node[right,yshift=-10] {$\eta \to -\infty$} -- (ar2);
    \end{tikzpicture}
    \end{gathered}
    \ee
    \item[(ii)] $\psi_n(\vp_1,\ldots,\vp_n)$ are the $n$-point wavefunction coefficients, which show both connected and disconnected components, and that receive contributions -- in principle -- at all loops: 
    \begin{equation}
    \psi_n = \sum_{\L \geq 0} \left[ \prod_{\ell=0}^{\L} \int \frac{\d^{d} \vell_e}{(2\pi)^{d}} \Psi^{(\ell)}(\vp_1, \ldots, \vp_{n}, \vell_1, \ldots, \vell_\L) \right]\,,
    \end{equation}
\end{itemize}
with
\begin{align}
\Psi^{(\ell)} & (\vp_1, \ldots, \vp_n , \vell_1, \ldots, \vell_{\L}) \big \vert_\text{ connected} = \delta^{(d)}(\vp_{1} + \ldots + \vp_n) \nonumber 
\begin{gathered}
\begin{tikzpicture}[scale=0.4]
    \coordinate (a) at (0.5,0);
    \coordinate (b) at (1.5,0);
    \coordinate (c) at (1,-1);
    \coordinate (d) at (2,-1);
    \coordinate (e) at (2.5,-0.5);
    \coordinate (f) at (2.5,-1);
    \coordinate (g) at (2.5,-1.5);
    \coordinate (i) at (3,-0.5);
    \coordinate (l) at (3.5,-0.5);
    \coordinate (m) at (4.5,-1.5);
    \coordinate (n) at (3.5,-1);
    \coordinate (o) at (3.5,-2);
    \coordinate (p) at (5.5,0);
    \coordinate (q) at (0,0);
    \coordinate (r) at (3,0);
    \coordinate (s) at (3.5,0);
    \coordinate (t) at (4.5,-0.5);
    \coordinate (u) at (4.5,0);
    \coordinate (v) at (5.5,-1);
    \coordinate (w) at (5,-2);
    \draw (q) to (p) node[right] {$\eta=0$};
    \draw (a) to (c);
    \draw (b) to (c);
    \draw (c) to (d);
    \draw (d) [out=90,in=180] to (e);
    \draw (d) [out=-90,in=180] to (g);
    \draw (e) to (g);
    \draw (e) to (i);
    \draw (i) to (r);
    \draw (i) to (l);
    \draw (l) to (s);
    \draw (f) to (n);
    \draw (g) to (o);
    \draw (o) to (m);
    \draw (n) to (m);
    \draw (n) to (t);
    \draw (t) to (u);
    \draw (t) to (v);
    \draw (v) to (m);
    \draw (v) to (w);
    \draw (o) [out=-20,in=210] to (w);
    \draw[thick,fill=black] (c) circle (1pt) (d) circle (1pt) (e) circle (1pt) (f) circle (1pt) (g) circle (1pt) (i) circle (1pt) (l) circle (1pt) (m) circle (1pt) (n) circle (1pt) (o) circle (1pt) (t) circle (1pt) (v) circle (1pt) (w) circle (1pt);
\end{tikzpicture}
\end{gathered}
\\ 
& \hspace{-1cm}= \delta^{(d)}(\vp_{1} + \ldots + \vp_n) \int_{-\infty}^{0} \prod_{s \in \mathcal{V}} \d\eta_{s} V(\eta_s) \left[ \prod_{j \in S} \phi_0(-E_j \eta_s) \right] \prod_{e \in \mathcal{E}} G(y_e;\eta_{s_e}\eta_{s'_{e}})\,, 
\end{align}
where $\mathcal{V}$ is the set of sites of the graph, while $\mathcal{E}$ is the set of edges. Furthermore,
\begin{itemize}
    \item $V(\eta_s)$ is the vertex function at the site $s$, which can depend on the warp factor of the metric -- here and in what follows we consider processes that occur in conformally flat backgrounds,
    \be \d s_{1+d}^2 = a^2(\eta) \big[- \d\eta^2 + \d\vx^2 \big], \qquad \eta \in \mathbb{R}_- \,.
    \ee
The vertex function can further depend on scalar products of momenta (in the case of derivative interactions -- for simplicity we will restrict to the case of polynomial interactions, unless otherwise specified).
\item
$\phi_0(-E_j \eta_s)$ is the classical solution stripped out of the boundary value $\Phi(\vp)$, e.g.,
\be \phi_{\text{cl}}(\vp , \eta) = \Phi(\vp) \phi_0(-E_j \eta_s).
\ee
Hence, $\phi_0(-E_j \eta_s)$ is taken to normalize to 1, that is, $\phi_0(-E_j \eta=0 ) = 1$.
\item 
$G(y_e; \eta_{s_{e}}, \eta_{s'_{e}})$ is the propagator that connects two sites $s_e$ and $s'_{e}$, with momentum $\vp_{e}$  flowing through it, and $y_e \equiv | \vp_e |$.
\end{itemize}

Because the fluctuations have to vanish at the boundary $\eta = 0$, the propagator has a three-term structure: two of them are the usual retarded and advanced contributions while the third one is purely a boundary term that manifestly breaks time-translation invariance:
\begin{align}
G(y_e;\eta_{s_e},\eta_{s'_{e}}) &= \frac{1}{2 \Re \big\{ \Psi_2(y_e) \big\}} \Big[ \bar{\phi}_0(-y_e \eta_{s_e}) \phi_0(-y_e \eta_{s'_{e}}) \Theta(\eta_{s_e} - \eta_{s'_{e}}) \\ 
&\qquad + \phi_0(-y_e \eta_{s_e}) \phi_0(-y_e \eta_{s'_{e}}) \Theta(\eta_{s'_{e}} - \eta_{s_e})  - \phi_0(- y_e\eta_{s_e}) \phi_0(-y_e \eta_{s'_{e}})\Big], \nonumber
\end{align}
where $\Re \big\{  \Psi_2(y_e) \big\}$ is the real part of the free two-point wavefunction.

\subsection{The wavefunction and its analytic properties}

Despite the fact that so far the discussion has been very general, it is nevertheless possible to extract important information on the expected analytic structure of the wavefunction.

First, consider a graph $\mathcal{G}$ that contributes to $\psi_n^{(\ell)}$. Following the Feynman rules discussed earlier, its contribution $\psi_{\mathcal{G}}$ can be written as
\begin{align}
\psi_{\mathcal{G}} & = \int_{-\infty}^0 \prod_{s \in \mathcal{V}} \left[ \d \eta_s V(\eta_s) \right] \prod_{j \in s}  \phi_0(-E_j \eta_s) \prod_{e \in \mathcal{E}} G(y_e; \eta_{s_e}, \eta_{s'_e})
\\ & = 
    \begin{gathered}
    \begin{tikzpicture}[baseline={([yshift=2ex]current bounding box.center)},scale=1.0001,thick]
    \coordinate (a) at (0,0);
    \coordinate (b) at (1,0);
    \coordinate (c) at (1.5,0);
    \coordinate (e) at (3,0);
    \coordinate (d) at (4,0);
    \coordinate (cen) at (2,-1);
    \draw[very thick] (a) -- (d);
    \draw[thick,fill=black] (b) circle (1pt) (c) circle (1pt) (e) circle (1pt);
    \draw[] (cen) to [out=180,in=-90] (b) node[above] {$\vp_1$};
    \draw[] (cen) to [out=180,in=-90] (c) node[above] {$\vp_2$};
    \draw[] (cen) to [out=0,in=-90] (e) node[above] {$\vp_n$};
    \draw[black!30,fill=black!30] (cen) circle (0.5);
    \end{tikzpicture}
    \end{gathered}
\end{align}
Let $\bar{\eta} = \eta_1 + \ldots + \eta_{n_s}$ be the ``center of mass'' time (with $n_s = \text{dim} \{ \mathcal{V} \}$) and $\eta_{s_e s'_e} = \frac{1}{2} (\eta_{s_e} - \eta_{s'_e})$.

Then, as $\bar{\eta} \to -\infty$, $\phi_0(-E_j\eta_s) \to f(\eta_s)\e^{iE_j \cdot \eta_s}$
because of the Bunch--Davies condition,
while $G(y_e;\eta_{s_e},\eta_{s'_e}) \to G_F(y_e; \eta_{s_e s'_e})\equiv G_R(y_e; \eta_{s_e s'_e}) + G_A(y_e; \eta_{s_e s'_e})$ where $G_F$, $G_R$, and $G_A$ correspond to the Feynman, retarded, and advanced propagators, respectively. That is, the propagator reduces to the Feynman propagator in this limit. Hence, 
\begin{align}
\psi_\mathcal{G} \to & \int_{-\infty} \d \bar{\eta} \, \tilde{V} (\bar{\eta}) \e^{i(E_1+ \ldots + E_{n}) \bar{\eta}}
\int_{-\infty}^{\infty}
\prod_{e\in\mathcal{E}} \d \eta_{s_e s'_e} \e^{i E_{s_e s'_e } \eta_{s_e s'_e}}
\prod_{e \in \mathcal{E}} G_F (y_e; \eta_{s_e s'_e}) \,.
\end{align}
Note that at a generic point in kinematic space, the integral is exponentially suppressed by $\e^{i(E_1 + \ldots + E_n) \bar{\eta}}$ upon regularization. However, for $E_1+ \ldots + E_n = 0$ the integral becomes singular with the type of singularity which depends on the function $\tilde{V}(\eta)$ and its coefficient given by the remaining integral which, for states that have a flat-space counterpart, is given by the high-energy limit of the graph contribution to the flat-space amplitude. Importantly, the locus $E_1+ \ldots + E_n = 0$ can be non-trivially reached if the energies are analytically continued such that a subset of them becomes negative -- such an analytic continuation maps these states to become ``out'' states.
Furthermore, as $\bar{\eta} \to -\infty$, the spacelike boundary becomes effectively infinitely far away, which can be seen from the fact that the propagator $G$ reduces to the Feynman propagator. This creates the setup for a flat-space scattering, with $E_{\text{tot}} = E_1+ \ldots + E_n = 0$  only having support on the total energy-conservation pole.

Note that the behavior just described is generic and does not depend on whether the states are massless or not: in all cases the mode functions reduce to the same exponential as a consequence of the Bunch--Davies condition in the far past. For example, in de Sitter space, massive states belong to its principal series and are characterized by a mode function given by a Hankel function with imaginary order parameter:
\be
\phi_0(- E \eta) = \sqrt{-E \eta} \, H^{(2)}_{i \lambda}(-E \eta) \,.
\ee
However, as $-E \eta \to \infty$, we get
\be \phi_0(-E \eta) \xrightarrow{\sim} \e^{i E \eta} \,.
\ee
In this sense, one obtains the high energy limit of a flat-space amplitude.

As a final comment, for states that do not have a flat-space counterpart, the analysis holds, with the important difference that now its coefficient no longer can be interpreted as the flat-space amplitude.

In summary, the Bunch--Davies wavefunction is singular on the locus $E_{\mathcal{G}} \equiv E_1 + \ldots + E_n = 0$ that lives outside the physical region and thus can be reached just upon analytic continuation.

The very same analysis can be carried out by considering the center-of-mass time associated to a given sub-process, identified by a subgraph $\mathfrak{g}$:
\begin{equation}
    \begin{gathered}
    \begin{tikzpicture}[baseline={([yshift=2ex]current bounding box.center)},scale=1.0001,thick]
    \coordinate (a) at (0,0);
    \coordinate (b) at (1,0);
    \coordinate (c) at (1.5,0);
    \coordinate (e) at (3,0);
    \coordinate (d) at (6,0);
    \coordinate (f) at (4.5,0);
    \coordinate (g) at (5,0);
    \coordinate (h) at (3.5,0);
    \coordinate (cen) at (2,-1);
    \coordinate (cen2) at (4,-1);
    \draw[very thick] (a) -- (d);
    \draw[] (cen) [in=-20,out=-160] to (cen2);
    \draw[thick,fill=black] (b) circle (1pt) (c) circle (1pt) (e) circle (1pt);
    \draw[] (cen) to [out=180,in=-90] (b);
    \draw[] (cen) to [out=180,in=-90] (c);
    \draw[] (cen) to [out=0,in=-90] (e);
    \draw[black!30,fill=black!30] (cen) circle (0.5);
    \draw[] (cen2) to [out=0,in=-90] (f);
    \draw[] (cen2) to [out=0,in=-90] (g);
    \draw[] (cen2) to [out=180,in=-90] (h);
    \draw[thick,fill=black] (f) circle (1pt) (g) circle (1pt) (h) circle (1pt);
    \coordinate (ar1) at (-0.5,-2);
    \coordinate (ar2) at (-0.5,-0.5);
    \draw[->, thick] (ar2) node[right,yshift=-10,xshift=-20] {$\bar{\eta}_L$} -- (ar1) node[below] {$-\infty$};
    \draw[black!30,fill=black!30] (cen2) circle (0.35);
    \draw[thick,Maroon,dashed] (cen) circle (0.8) node[below,yshift=-25,scale=0.8] {$\mathfrak{g}$}; 
    \draw[thick,RoyalBlue,dashed] (cen2) circle (0.6) node[below,yshift=-20,scale=0.8] {$\bar{\mathfrak{g}}$}; 
    \end{tikzpicture}
    \end{gathered}
\end{equation}
In this case the singularity is located at
\be
E_{\mathfrak{g}} \equiv \sum_{s \in \mathcal{V}_{\mathfrak{g}}} \sum_{j \in s} E_j + \sum_{c \in \mathcal{E}_{\mathfrak{g}}^{\text{ext}}} y_e = 0\,,
\ee
where $\mathcal{V}_{\mathfrak{g}}$ is the set of sites in $\mathfrak{g}$, while $\mathcal{E}_{\mathfrak{g}}^{\text{ext}}$ is the set of edges departing from $\mathfrak{g}$. On this hyperplane, the wavefunction coefficient factorizes into a lower-point scattering amplitude and a linear combination of lower point wavefunctions: 
\begin{equation}
\tilde{\psi}_{\mathcal{G}} \xrightarrow[]{E_{\mathfrak{g}} \to 0} A_{\mathfrak{g}} \sum_{\sigma_{\slashed{e}} =   \pm 1} \frac{1}{ 2 \Re \big\{ \Psi_2(y_{\slashed{e}}) \big\}} \Psi_{\bar{\mathfrak{g}}}(\slashed{\mathcal{E}} (\sigma_{\slashed{e}}),R)\,,
\label{eq:psitilde}
\end{equation}
where $A_{\mathfrak{g}}$ is the contribution to the flat-space amplitude associated to the subgraph $\mathfrak{g}$, while $\Psi_{\bar{\mathfrak{g}}}(\slashed{\mathcal{E}},R)$ is the wavefunction associated with $\bar{\mathfrak{g}}$ whose external energies contain the energies of the edges $\slashed{\mathcal{E}}$ which connect it to $\mathfrak{g}$ each of which is taken with the sign $\sigma_{\slashed{e}}$. This can be understood thinking that in this limit just one of the  time-ordered terms in the propagator $\mathfrak{g}$ survives, {e.g.,}
\begin{equation}
\hspace{-0.5cm}
G(y_{\slashed{e}}; \eta_{s_{\slashed{e}}}, \eta_{s'_{\slashed{e}}}) \rightarrow G_{\text{ret}} {-} G_{\partial} = \frac{1}{2 \Re \big\{ \Psi_2(y_{\slashed{e}}) \big\}} \phi(-y \eta_{s'_{\slashed{e}}} )\left[\bar{\phi}_0(-y \eta_{s_{\slashed{e}}}) {-} \phi_0 (- y \eta_{s_{\slashed{e}}}) \right],
\end{equation}
giving rise to the sum in the right-hand side of~\eqref{eq:psitilde}.

In summary, the wavefunction shows singularities in correspondence with the vanishing loci of the energies associated to its sub-processes, $\{ E_{\mathfrak{g}} = 0, \quad\mathfrak{g}\subseteq \mathcal{G} \}$. The coefficient of each of these singularities is the factorization into a lower-point amplitude and a linear combination of lower-point wavefunctions. 

Hence, given a graph $\mathcal{G}$, there is a one-to-one correspondence between the singularities of $\psi_{\mathcal{G}}$ and the subgraphs $\{\mathfrak{g}\subseteq \mathcal{G} \}$.

\subsection{Cosmological integrals}

Let us consider a large class of scalar toy models consisting of polynomial interactions with time-dependent couplings:
\begin{equation}
S[\phi] = - \int_{-\infty}^0 \d \eta \int \d^d \vx \left[\frac{1}{2}(\partial \phi)^2 - \sum_{k \geq 3}\frac{\lambda_k(\eta)}{k!}\phi^k \right] \,.
\end{equation}
This class contains as a special case conformally-coupled scalars in FRW-cosmologies, provided that $\lambda_k(\eta)$ is identified to be 
\begin{equation}
\lambda_k (\eta ) : = \lambda \left[ a(\eta) \right]^{2 - \frac{(k-2)(d-1)}{2}} \,,
\end{equation} 
where $a(\eta)$ is the warp factor of the FRW metric $\d s^2 = a^2(\eta)[- \d \eta^2 + \d \vx^2] $.

Furthermore, the conformally-coupled scalar can be mapped to other states via a differential operator. Then, given a graph $\mathcal{G}$, the wavefunction coefficient associated to it is given by 
\begin{equation}
\tilde{\psi}_{\mathcal{G}} = i^{n_s}\int_{-\infty}^0 \prod_{s \in \mathcal{V}} \d \eta_s \lambda_k(\eta_s)\e^{i X_{s} \eta_s} \prod_{e \in \mathcal{E}} G(y_e; \eta_{s_e}, \eta_{s'_e})\,,
\end{equation}
where $X_{s} = \sum_{j \in s} E_j$ is the total energy of the states in the graph $\mathcal{G}$, and $n_s$ is the number of external states in $\mathcal{G}$. The mode functions are simple exponentials, and thus the propagators are,
\begin{equation}
\begin{split}
    G(y_e; \eta_{s_e}, \eta_{s'_e}) = \frac{1}{2y_e} \Big[&\e^{-i y_e (\eta_{s_e} - \eta_{s'_e})} \Theta(\eta_{s_e} - \eta_{s'_e}) +\e^{i y_e (\eta_{s_e} - \eta_{s'_e})} \Theta(\eta_{s'_e} - \eta_{s_e}) \\& -\e^{i y_e (\eta_{s_e} + \eta_{s'_e})} \Big]\,.
\end{split}
\end{equation}
Finally, the spatial momentum-conserving delta function has been stripped off.

In order to perform the time integrals over $\eta$, it naively seems to be necessary to specify $\lambda(\eta)$ at this stage, which encodes the details of the cosmology. However, we can still extract more information while staying agnostic to the details of the cosmology. To that end, we consider the following integral expression for $\lambda(\eta)$:
\begin{equation}
\lambda(\eta) : = i \int_{-\infty}^{\infty} \d z \,\e^{i z \eta} \tilde{\lambda}(z) = \int_{0}^{\infty} \d z \, \e^{i z \eta} \tilde{\lambda}(z),
\end{equation}
with the second equality coming from the fact that $\lambda(\eta)$ has support only on $\Theta(-\eta)$ as $\eta\in\mathbb{R}_-$.
Hence,
\begin{align}
\tilde{\psi}_{\mathcal{G}} &= \int_{0}^{\infty} \prod_{s \in \mathcal{V}} \d z_s \, \tilde{\lambda}(z_s) i^{n_s}
\int_{-\infty}^0 \prod_{s \in \mathcal{V}} d \eta_s \, 
e^{i (X_{s}+z_s) \eta_s} \prod_{e \in \mathcal{E}} G(y_e; \eta_{s_e}, \eta_{s'_e}) \\
&= \prod_{s \in \mathcal{V}} \int_{X_s}^{\infty} \d x_s \, \tilde{\lambda} (x_s - X_s) i^{n_s}  \int_{-\infty}^0 \prod_{s \in \mathcal{V}} \d \eta_s\e^{i x_s \eta_s} \prod_{e \in \mathcal{E}} G(y_e; \eta_{s_e}, \eta_{s'_e}) 
\\
&= \prod_{s \in \mathcal{V}} \int_{X_s}^{\infty} \d x_s \tilde{\lambda} (x_s - X_s) \psi_{\mathcal{G}}(x,y),
\end{align}
where the second line comes from the simple change of variables $x_s = X_{s} + z_s$. Furthermore, recall that
\begin{equation}
\psi_{\mathcal{G}}(x,y) \equiv  i^{n_s} \int_{-\infty}^{0} \prod_{s \in \mathcal{V}} \d \eta_s\e^{i x_s \eta_s} \prod_{e \in \mathcal{E}} G(y_e; \eta_{s_e}, \eta_{s'_e}),
\end{equation}
is nothing but the flat-space wavefunction, with $x \equiv \{ x_s \} _{s \in \mathcal{V}}$ and $\{y_e\}_{e \in \mathcal{E}}$ parametrising the flat-space kinematic space.

Hence, the formula
\begin{equation}
    \tilde{\psi}_{\mathcal{G}} (X,y) = \prod_{s \in \mathcal{V}} \int_{X_s}^\infty \d x_s \, \tilde{\lambda} (x_s-X_s) \psi_{\mathcal{G}} (x,y)\,,
\end{equation}
can be seen as a map between the flat-space wavefunction and the FRW one, with the details of the cosmology encoded in the integration measure $\tilde{\lambda}(x_s-X_s)$. This formula makes explicit that part of the structure of $\tilde{\psi}_{\mathcal{G}} (X,y)$ is determined by $\psi_{\mathcal{G}} (x,y)$ -- in particular the loci where a subset of energies vanishes, $\{ E_{\mathfrak{g}} (x,y) = 0, \; \mathfrak{g} \subseteq \mathcal{G}\}$, are mapped into the loci $\{ E_{\mathfrak{g}} (X,y) = 0, \; \mathfrak{g} \subseteq \mathcal{G}\}$. The full singularity structure of $\tilde{\psi}_{\mathcal{G}}$ is further constrained by the way in which the measure of integration $\tilde{\lambda}$, the integrand $\psi_{\mathcal{G}} (x,y)$, and the integration contour over $x_s$ relate to each other.

For power-law cosmologies with $a (\eta) = \big( - \frac{\ell}{\eta} \big)^\gamma$ ($\Re(\gamma)>0$), the map is simply a Mellin transform
\begin{align}
    \tilde{\psi}_{\mathcal{G}} (X,y) & = \frac{i^\alpha}{\Gamma(\alpha)} \ell^\alpha \prod_{s \in \mathcal{V}} \int_{X_s}^\infty \d x_s (x_s-X_s)^{\alpha-1} \psi_{\mathcal{G}} (x,y)
    \\ & = 
    \frac{i^{\alpha}}{\Gamma(\alpha)} \ell^\alpha \int_0^\infty \prod_{s \in \mathcal{V} } \frac{\d x_s}{x_s} x_s^\alpha  \, \psi_{\mathcal{G}} (x+X,y) \,.
\end{align}
As is customary in the literature on scattering amplitudes, it is possible to study the integrand $\psi_\mathcal{G}(x,y)$ and later extend the information on the integrated function. We stress once more that the integrand is the same for each cosmology: its structure will affect the one of $\tilde{\psi}_{\mathcal{G}}(X,y)$ for any cosmology.

First, note that the wavefunction coefficient $\psi_{\mathcal{G}}(x,y)$ associated with a graph $\mathcal{G}$ depends only on the labels associated with the sites of $\mathcal{G}$, $\{x_s, s \in \mathcal{V} \}$, and on its edges $\{y_e, e \in \mathcal{E} \}$. Hence, rather than considering the full graph, it is possible to define a reduced graph that associates a weight to each site, e.g.,
\begin{equation}
    \begin{gathered}
    \begin{tikzpicture}[baseline={([yshift=2ex]current bounding box.center)},scale=1.0001]
    \coordinate (a) at (0,0);
    \coordinate (b) at (0.5,0);
    \coordinate (c) at (1.5,0);
    \coordinate (d) at (2.5,0);
    \coordinate (e) at (3,0);
    \coordinate (la) at (0.5,-1);
    \coordinate (lb) at (1.5,-1);
    \coordinate (lc) at (2.5,-1);
    \draw[thick] (a) -- (e) (a) --++ (180:0.2) (e) --++ (0:0.2);
    \draw[thick,fill=black] (a) circle (1pt) node[above, scale=1.0001] {$E_1$} (b) circle (1pt) node[above, scale=1.0001] {$E_2$} (c) circle (1pt) node[above, scale=1.0001] {$E_3$} (d) circle (1pt) node[above, scale=1.0001] {$E_4$} (e) circle (1pt) node[above, scale=1.0001] {$E_5$};
    \draw[thick,fill=black] (la) circle (1pt) (lb) circle (1pt)  (lc) circle (1pt) ;
    \draw[thick] (a) -- (la) (b) -- (la) (c) -- (lb) (d) -- (lc) (e) -- (lc) (la) -- (lc);
    \node[scale=1.0001] at (1,-1.3) {$y_{12}$};
    \node[scale=1.0001] at (2,-1.3) {$y_{23}$};
    \end{tikzpicture}
    \end{gathered}
    \longrightarrow
    \begin{gathered}
    \begin{tikzpicture}[baseline={([yshift=2ex]current bounding box.center)},scale=0.8]
    \coordinate (a) at (0,0);
    \coordinate (b) at (1.5,0);
    \coordinate (c) at (3,0);
    \draw[thick] (a) -- (b);
    \draw[thick] (b) -- (c);
    \draw[thick,fill=black] (a) circle (2pt) node[below] {$x_1$};
    \draw[thick,fill=black] (b) circle (2pt) node[below] {$x_2$};
    \draw[thick,fill=black] (c) circle (2pt) node[below] {$x_3$};
    \node[] at (0.75,0.35) {$y_{12}$};
    \node[] at (2.25,0.35) {$y_{23}$};
    \end{tikzpicture}
    \end{gathered}
\label{eq:reduced_graph}
\end{equation}
with $E_j = |\vp_j|$, $y_{12} = |\vp_1+\vp_2|$ and $y_{23} = |\vp_2+\vp_3|$.
The reduced graph on the right-hand side is simply obtained by suppressing the external legs and instead weighting each site $s$ with the external energies $x_j = \sum_{j \in s} |\vp_s|$. In particular, the weighted reduced graph on the right-hand side of~\eqref{eq:reduced_graph} has $x_1 = E_1+E_2$, $x_2 = E_3$ and $x_3 = E_4+E_5$. The integration over $\{x_s, s \in \mathcal{V}\}$ then maps the flat-space kinematics onto the FRW one. Then, the reduced weighted graph represents the time integration in flat space. Direct integration returns an expression with $3^{n_e}$ terms ($n_e = \text{dim} \{\mathcal{E} \}$) with spurious poles at $\{y_e=0,  \; e \in \mathcal{E}\}$ because of the $3$-term structure of $G$, e.g.,
\begin{align}
    \begin{gathered}
    \begin{tikzpicture}[baseline={([yshift=2ex]current bounding box.center)}]
    \coordinate (a) at (0,0);
    \coordinate (b) at (1.5,0);
    \draw[thick] (a) -- (b);
    \draw[thick,fill=black] (a) circle (2pt) node[below] {$x_1$};
    \draw[thick,fill=black] (b) circle (2pt) node[below] {$x_2$};
    \node[] at (0.75,0.35) {$y_{12}$};
    \end{tikzpicture}
    \end{gathered}
& =
i^2 \int_{-\infty}^0 \d \eta_1\e^{i x_1 \eta_1} \int_{-\infty}^0 \d \eta_2\e^{i x_2 \eta_2} G(y_{12}; \eta_1, \eta_2)
\nonumber
\\
& = 
\underbracket[0.4pt]{{}\frac{1}{2 y_{12} (x_1+x_2) (x_2+ y_{12})} }_{\text{retarded contribution}}
+
\underbracket[0.4pt]{{}\frac{1}{2 y_{12} (x_1+x_2) (x_1+ y_{12})} }_{\text{advanced contribution}}
-
\underbracket[0.4pt]{{}\frac{1}{2 y_{12} (x_1+y_{12}) (y_{12} + x_2)} }_{\text{boundary term}}
\nonumber
\\ & = 
\frac{1}{(x_1+x_2)(x_1+y)(y+x_2)}\,.
\end{align}
The last line explicitly shows that
\begin{itemize}
    \item[(i)] $y_{12}=0$ is a spurious pole,
    \item[(ii)] the graph has only singularities corresponding to subgraphs, as predicted by the general argument discussed earlier
    \begin{equation}
    \begin{gathered}
    \begin{tikzpicture}[baseline={([yshift=2ex]current bounding box.center)}]
    \coordinate (a) at (0,0);
    \coordinate (b) at (1.5,0);
    \draw[thick] (a) -- (b);
    \draw[thick,fill=black] (a) circle (2pt) node[below,yshift=-4] {$x_1$};
    \draw[thick,fill=black] (b) circle (2pt) node[below,yshift=-4] {$x_2$};
    \node[] at (0.75,0.35) {$y_{12}$};
    \draw[Maroon, thick] (0.75, 0) ellipse (0.95cm and 0.2cm);
    \node[scale=0.8] at (0.75,-1) {$E_{\mathfrak{g}=\mathcal{G}} \equiv E_{\mathcal{G}} = x_1+x_2$};
    \end{tikzpicture}
    \hspace{0.9cm}
    \begin{tikzpicture}[baseline={([yshift=2ex]current bounding box.center)}]
    \coordinate (a) at (0,0);
    \coordinate (b) at (1.5,0);
    \draw[thick] (a) -- (b);
    \draw[thick,fill=black] (a) circle (2pt) node[below,yshift=-4] {$x_1$} node[above,yshift=4] {$\mathfrak{g}_1$};
    \draw[thick,fill=black] (b) circle (2pt) node[below,yshift=-4] {$x_2$};
    \node[] at (0.75,0.35) {$y_{12}$};
    \draw[Maroon, thick] (a) ellipse (0.2cm and 0.2cm);
    \node[scale=0.8] at (0.75,-1) {$E_{\mathfrak{g}=\mathfrak{g}_1} \equiv E_{\mathfrak{g}_1} = x_1+y_{12}$};
    \end{tikzpicture}
    \hspace{0.9cm}
    \begin{tikzpicture}[baseline={([yshift=2ex]current bounding box.center)}]
    \coordinate (a) at (0,0);
    \coordinate (b) at (1.5,0);
    \draw[thick] (a) -- (b);
    \draw[thick,fill=black] (a) circle (2pt) node[below,yshift=-4] {$x_1$};
    \draw[thick,fill=black] (b) circle (2pt) node[below,yshift=-4] {$x_2$} node[above,yshift=4] {$\mathfrak{g}_2$};
    \node[] at (0.75,0.35) {$y_{12}$};
    \draw[Maroon, thick] (b) ellipse (0.2cm and 0.2cm);
    \node[scale=0.8] at (0.75,-1) {$E_{\mathfrak{g}=\mathfrak{g}_2} \equiv E_{\mathfrak{g}_2} = y_{12} + x_2$};
    \end{tikzpicture}
    \end{gathered}
    \end{equation}
    with the general formula $E_{\mathfrak{g}} = \sum_{s \in \mathcal{V}_s}x_s + \sum_{e \in \mathcal{E}_{\mathfrak{g}}^{\text{ext}}} y_e$, where $\mathcal{V}_s$ is the set of sites in $\mathfrak{g}$, while $\mathcal{E}^{\text{ext}}_{\mathfrak{g}}$ is the set of edges departing from $\mathfrak{g}$,
    \item[(iii)] the expression obtained via Feynman rules is redundant and can be simplified in one which contains just physical singularities.
\end{itemize}
A general question is whether such a formula can be obtained, bypassing the Feynman rules, and whether there is a mathematical structure behind it.

Let us consider a general graph $\mathcal{G}$ and the time integral associated to it
\begin{equation}
    \psi_\mathcal{G} = i^{n_s} \int_{-\infty}^0 \prod_{s \in \mathcal{V}} \left[ \d \eta_s\e^{i x_s \eta_s} \right]
    \prod_{e \in \mathcal{E}} G(y_e ; \eta_{s_e}, \eta_{s_e'}) \,.
\end{equation}
Now let us consider the very same integral but with the total time translation generator acting on the full integrand -- this integral is zero because of the boundary conditions:
\begin{equation}
    0 = i^{n_s} \int_{-\infty}^0 \prod_{s \in \mathcal{V}} \left[ \d \eta_s \right] \hat{\Delta} \left\{ \prod_{s \in \mathcal{V}}\e^{i x_s \eta_s} \prod_{e \in \mathcal{E}}  G(y_e ; \eta_{s_e}, \eta_{s_e'})  \right\} \,,
\end{equation}
with $\hat{\Delta} = - i \sum_{s \in \mathcal{V}} \partial_{ \eta_s }$. However, we can make $\hat{\Delta}$ act on the external states $\e^{i x_s \eta_s}$ and then on the propagators $G(y_e ; \eta_{s_e}, \eta_{s_e'})$,
\begin{align}
    0 & = i^{n_s} \int_{-\infty}^0 \prod_{s \in \mathcal{V}} \left[ \d \eta_s \right] \hat{\Delta} \Big[ \prod_{s \in \mathcal{V}}\e^{i x_s \eta_s}\Big] \prod_{e \in \mathcal{E}} G(y_e ; \eta_{s_e}, \eta_{s_e'})
    \\ & \hspace{2cm}
    +
    i^{n_s} \int_{-\infty}^0 \prod_{s \in \mathcal{V}} \left[ \d \eta_s\e^{i x_s \eta_s} \right] \hat{\Delta} \Big[ \prod_{e \in \mathcal{E}} G(y_e ; \eta_{s_e}, \eta_{s_e'}) \Big]
    \\ & = 
    \Big( \sum_{s \in \mathcal{V} x_s} x_s \Big) \psi_{\mathcal{G}} - \sum_{e \in \mathcal{E}} \psi_{\mathcal{G}\setminus \{e\}} (x_j^{(e)},g)\,,
\end{align}
where $\mathcal{G} \setminus \{e\}$ represents the graph obtained by deleting the edge, while $x_s^{(e)}: = x_s + \sum_{\{e\} \cap \mathcal{E}_s}y_e$,  $\mathcal{E}_s$ being the set of edges departing from $s$. The relation above provides a recursion relation. Note that $\mathcal{G} \setminus \{e \}$ can be either disconnected or connected. Separating these two classes of contributions then results in
\begin{equation}
    \Big( \sum_{s \in \mathcal{V} x_s} x_s \Big) \psi_{\mathcal{G}} = \sum_{e \in \mathcal{E}} 
    \underbracket[0.4pt]{{}
    \begin{gathered}
    \begin{tikzpicture}[baseline={([yshift=0ex]current bounding box.center)}]
    \coordinate (a) at (0,0);
    \coordinate (b) at (1.5,0);
    \filldraw[fill=black!20!white, draw=black, thick] (a) circle (0.25 cm);
    \draw[dashed] ($(a)+(0.25,0)$) -- ($(b)-(0.25,0)$);
    \filldraw[fill=black!20!white, draw=black, thick] (b) circle (0.25 cm);
    \draw[thick,fill=black] ($(a)+(0.25,0)$) circle (1pt) node[below, scale=0.6, xshift=8, yshift=-8] {$x_{s_e}{+}y_e$};
    \draw[thick,fill=black] ($(b)-(0.25,0)$) circle (1pt) node[below, scale=0.6, xshift=-14, yshift=24] {$x_{s_e'}{+}y_e$};
    \end{tikzpicture}
    \end{gathered}
    }_{\text{disconnected terms}}
    +
    \sum_{e \in \mathcal{E}} 
    \underbracket[0.4pt]{{}
    \begin{gathered}
    \begin{tikzpicture}
    \coordinate (a) at (0,0);
    \coordinate (b) at (0,-0.25);
    \filldraw[fill=white, draw=black, thick] (b) circle (0.25 cm);
    \filldraw[fill=black!20!white, draw=black, thick] (a) circle (0.25 cm);
    \draw[thick,fill=black] (-0.216506,-0.125) circle (1pt) node[right, scale=0.6, xshift=-42] {$x_{s_e}{+}y_e$};
    \draw[thick,fill=black] (0.216506,-0.125) circle (1pt) node[right, scale=0.6] {$x_{s_e'}{+}y_e$};
    \end{tikzpicture}
    \end{gathered}
    }_{\text{connected terms}}\,,
\end{equation}
with the dashed line indicating the deleted edge. Note that the term $\big( \sum_{s \in \mathcal{V}} x_s \big) \psi_{\mathcal{G}} (x,y)$ has been obtained because
\begin{equation}
    \hat{\Delta} \prod_{s \in \mathcal{V}}\e^{i x_s \eta_s} = \Big( \sum_{s \in \mathcal{V}} x_s\Big) \prod_{s \in \mathcal{V}}\e^{i x_s \eta_s} \,,
\end{equation}
the term $\sum_{e \in \mathcal{E}} \psi_{\mathcal{G}\setminus \{e\}} (x_j^{(e)},y)$ instead uses the relation $\hat{\Delta} G = -\e^{i y_e (\eta_{s_e}+ \eta_{s_e'})}$.

The resulting recursion relation can be translated into simple graphical rules, by iteratively considering all possible subgraphs and summing all possible ways in which subgraphs can be obtained by erasing one edge, while associated to each subgraph the inverse of its total energy $1/E_{\mathfrak{g}}$, $E_{\mathfrak{g}} : = \sum_{s \in \mathcal{V}_{\mathfrak{g}}} x_s + \sum_{e \in \mathcal{E}_{\mathfrak{g}}^{\text{ext}}}y_e$.
\begin{mdexample}
We first look at the simplest example with only two sites. The first subgraph is simply the graph itself:
    \begin{equation}
    \begin{gathered}
    \begin{tikzpicture}[baseline={([yshift=2ex]current bounding box.center)}]
    \coordinate (a) at (0,0);
    \coordinate (b) at (1.5,0);
    \draw[thick] (a) -- (b);
    \draw[thick,fill=black] (a) circle (2pt) node[below,yshift=-4] {$x_1$};
    \draw[thick,fill=black] (b) circle (2pt) node[below,yshift=-4] {$x_2$};
    \node[] at (0.75,0.35) {$y_{12}$};
    \coordinate (c) at (2.5,0);
    \draw[->,thick] (c) --++ (1,0);
    \node[scale=0.8] at (3.0,-0.7) {$\substack{\text{the first subgraph} \\ \text{is the graph itself}}$};
    \begin{scope}[shift={(5,0)}]
    \coordinate (a) at (0,0);
    \coordinate (b) at (1.5,0);
    \draw[thick] (a) -- (b);
    \draw[thick,fill=black] (a) circle (2pt);
    \draw[thick,fill=black] (b) circle (2pt);
    \draw[Maroon, thick] (0.75, 0) ellipse (0.95cm and 0.2cm);
    \coordinate (c) at (2.5,0);
    \draw[->,thick] (c) --++ (1,0);
    \node[scale=0.8] at (3.0,-0.7) {$\substack{\text{there is just one} \\ \text{way in which one} \\ \text{can generate a subgraph}
    \\ \text{erasing an edge}
    }$};
    \node[scale=0.8] at (0.7,-0.7) {$\frac{1}{x_1+x_2}$};
    \end{scope}
    \begin{scope}[shift={(10,0)}]
    \coordinate (a) at (0,0);
    \coordinate (b) at (1.5,0);
    \node[scale=0.8] at (0.7,-0.7) {$\frac{1}{x_1+x_2} \frac{1}{x_1+y_{12}} \frac{1}{y_{12} + x_2}$};
    \draw[thick] (a) -- (b);
    \draw[thick,fill=black] (a) circle (2pt);
    \draw[thick,fill=black] (b) circle (2pt);
    \draw[Maroon, thick] (0.75, 0) ellipse (1.1cm and 0.3cm);
    \draw[Maroon, thick] (b) ellipse (0.15cm and 0.15cm);
    \draw[Maroon, thick] (a) ellipse (0.15cm and 0.15cm);
    \end{scope}
    \end{tikzpicture}
    \end{gathered}
\end{equation}
\end{mdexample}

\begin{mdexample}
Next, we look at an example with three sites. The first encircling is the one of all the sites:
\begin{equation}
    \begin{gathered}
    \begin{tikzpicture}[baseline={([yshift=2ex]current bounding box.center)}]
    \coordinate (a) at (0,0);
    \coordinate (b) at (1.5,0);
    \coordinate (c) at (3,0);
    \draw[thick] (a) -- (b) -- (c);
    \draw[thick,fill=black] (a) circle (2pt) node[below,yshift=-4] {$x_1$}
    (b) circle (2pt) node[below,yshift=-4] {$x_2$}
    (c) circle (2pt) node[below,yshift=-4] {$x_3$}
    ;
    \node[] at (0.75,0.35) {$y_{12}$};
    \node[] at (2.25,0.35) {$y_{23}$};
    \coordinate (d) at (3.5,0);
    \draw[->,thick] (d) --++ (1,0);
    \begin{scope}[shift={(5,0)}]
    \coordinate (a) at (0,0);
    \coordinate (b) at (1.5,0);
    \coordinate (c) at (3,0);
    \draw[thick] (a) -- (b) -- (c);
    \draw[thick,fill=black] (a) circle (2pt)  (b) circle (2pt) (c) circle (2pt);
    \draw[Maroon, thick] (b) ellipse (1.8cm and 0.2cm);
    \node[scale=0.8] at (1.5,-0.5) {$\frac{1}{x_1+x_2+x_3}$};
    \end{scope}
    \end{tikzpicture}
    \end{gathered}
\end{equation}
There are now two different ways of generating subgraphs by deleting an edge. In both cases, there is only one way to generate further subgraphs by deleting one edge. The first possibility is,
\begin{equation}
    \begin{gathered}
    \begin{tikzpicture}[baseline={([yshift=2ex]current bounding box.center)}]
    \coordinate (a) at (0,0);
    \coordinate (b) at (1.5,0);
    \coordinate (c) at (3,0);
    \draw[thick] (a) -- (b) -- (c);
    \draw[thick,fill=black] (a) circle (2pt)  (b) circle (2pt) (c) circle (2pt);
    \draw[Maroon, thick] (b) ellipse (1.8cm and 0.2cm);
    \node[scale=0.8] at (1.5,-0.9) {$\frac{1}{x_1+x_2+x_3}$};
    \coordinate (d) at (3.5,0);
    \draw[->,thick] (d) --++ (1,0);
    \begin{scope}[shift={(5,0)}]
    \coordinate (a) at (0,0);
    \coordinate (b) at (1.5,0);
    \coordinate (c) at (3,0);
    \draw[thick] (a) -- (b) -- (c);
    \draw[thick,fill=black] (a) circle (2pt)  (b) circle (2pt) (c) circle (2pt);
    \draw[Maroon, thick] (b) ellipse (1.8cm and 0.4cm);
    \draw[Maroon, thick] (0.75,0) ellipse (1cm and 0.25cm);
    \draw[Maroon, thick] (c) ellipse (0.15cm and 0.15cm);
    \node[scale=0.8] at (1.5,-0.9) {$\frac{1}{x_1+x_2+x_3} \frac{1}{x_1+x_2+y_{23}} \frac{1}{y_{23} + x_3}$};
    \coordinate (d) at (3.5,0);
    \draw[->,thick] (d) --++ (1,0);
    \end{scope}
    \begin{scope}[shift={(10,0)}]
    \coordinate (a) at (0,0);
    \coordinate (b) at (1.5,0);
    \coordinate (c) at (3,0);
    \draw[thick] (a) -- (b) -- (c);
    \draw[thick,fill=black] (a) circle (2pt)  (b) circle (2pt) (c) circle (2pt);
    \draw[Maroon, thick] (b) ellipse (1.8cm and 0.4cm);
    \draw[Maroon, thick] (0.75,0) ellipse (1cm and 0.25cm);
    \draw[Maroon, thick] (c) ellipse (0.12cm and 0.12cm);
    \draw[Maroon, thick] (a) ellipse (0.12cm and 0.12cm);
    \draw[Maroon, thick] (b) ellipse (0.12cm and 0.12cm);
    \node[scale=0.8] at (1.5,-0.9) {$\substack{\frac{1}{x_1+x_2+x_3} \frac{1}{x_1+x_2+y_{23}} \frac{1}{y_{23} + x_3} \\[3pt] 
    \hspace{0.5cm} \times \frac{1}{x_1+y_{12}} \frac{1}{y_{12} + x_2 + y_{23}}}$};
    \end{scope}
    \end{tikzpicture}
    \end{gathered}
\end{equation}
while the second one is
\begin{equation}
    \begin{gathered}
    \begin{tikzpicture}[baseline={([yshift=2ex]current bounding box.center)}]
    \coordinate (a) at (0,0);
    \coordinate (b) at (1.5,0);
    \coordinate (c) at (3,0);
    \draw[thick] (a) -- (b) -- (c);
    \draw[thick,fill=black] (a) circle (2pt)  (b) circle (2pt) (c) circle (2pt);
    \draw[Maroon, thick] (b) ellipse (1.8cm and 0.2cm);
    \node[scale=0.8] at (1.5,-0.9) {$\frac{1}{x_1+x_2+x_3}$};
    \coordinate (d) at (3.5,0);
    \draw[->,thick] (d) --++ (1,0);
    \begin{scope}[shift={(5,0)}]
    \coordinate (a) at (0,0);
    \coordinate (b) at (1.5,0);
    \coordinate (c) at (3,0);
    \draw[thick] (a) -- (b) -- (c);
    \draw[thick,fill=black] (a) circle (2pt)  (b) circle (2pt) (c) circle (2pt);
    \draw[Maroon, thick] (b) ellipse (1.8cm and 0.4cm);
    \draw[Maroon, thick] (2.25,0) ellipse (1cm and 0.25cm);
    \draw[Maroon, thick] (a) ellipse (0.15cm and 0.15cm);
    \node[scale=0.8] at (1.5,-0.9) {$\frac{1}{x_1+x_2+x_3} \frac{1}{x_1+y_{12}} \frac{1}{y_{12}+x_2+x_3}$};
    \coordinate (d) at (3.5,0);
    \draw[->,thick] (d) --++ (1,0);
    \end{scope}
    \begin{scope}[shift={(10,0)}]
    \coordinate (a) at (0,0);
    \coordinate (b) at (1.5,0);
    \coordinate (c) at (3,0);
    \draw[thick] (a) -- (b) -- (c);
    \draw[thick,fill=black] (a) circle (2pt)  (b) circle (2pt) (c) circle (2pt);
    \draw[Maroon, thick] (b) ellipse (1.8cm and 0.4cm);
    \draw[Maroon, thick] (2.25,0) ellipse (1cm and 0.25cm);
    \draw[Maroon, thick] (c) ellipse (0.12cm and 0.12cm);
    \draw[Maroon, thick] (a) ellipse (0.12cm and 0.12cm);
    \draw[Maroon, thick] (b) ellipse (0.12cm and 0.12cm);
    \node[scale=0.8] at (1.5,-0.9) {$\substack{\frac{1}{x_1+x_2+x_3} \frac{1}{x_1+y_{12}} \frac{1}{y_{12} + x_2+x_3} \\[3pt] 
    \hspace{0.5cm} \times \frac{1}{y_{12} + x_2 + y_{23}} \frac{1}{y_{23}+x_3}}$};
    \end{scope}
    \end{tikzpicture}
    \end{gathered}
\end{equation}
The final result is the sum of such contributions:
\begin{align}
    \begin{gathered}
    \begin{tikzpicture}[baseline={([yshift=2ex]current bounding box.center)}]
    \coordinate (a) at (0,0);
    \coordinate (b) at (1.5,0);
    \coordinate (c) at (3,0);
    \draw[thick] (a) -- (b) -- (c);
    \draw[thick,fill=black] (a) circle (2pt) node[below,yshift=-4] {$x_1$}
    (b) circle (2pt) node[below,yshift=-4] {$x_2$}
    (c) circle (2pt) node[below,yshift=-4] {$x_3$}
    ;
    \node[] at (0.75,0.35) {$y_{12}$};
    \node[] at (2.25,0.35) {$y_{23}$};
\end{tikzpicture}
\end{gathered}
& = 
\frac{1}{(x_1+x_2+x_3) (x_1+y_{12}) (y_{12}+x_2+y_{13})(y_{23}+x_3)}
\nonumber
\\ & \hspace{3cm} \times \left[ \frac{1}{x_1+x_2+y_{23}} + \frac{1}{y_{12}+x_2+x_3} \right] \,.
\end{align}
\end{mdexample}

\begin{mdexample}
    As a last example, let us consider the 2-site one-loop graph,
    \begin{equation}
    \begin{gathered}
    \begin{tikzpicture}[baseline={([yshift=2ex]current bounding box.center)},scale=0.8]
    \coordinate (a) at (0,0);
    \coordinate (b) at (1.5,0);
    \draw[thick] (0.75,0) circle (0.75);
    \draw[thick,fill=black] (a) circle (2pt) node[left] {$x_1$};
    \draw[thick,fill=black] (b) circle (2pt) node[right] {$x_2$};
    \node[] at (0.75,1) {$y_{12}$};
    \node[] at (0.75,-1) {$y_{21}$};
    \end{tikzpicture}
    \end{gathered}
    \end{equation}
    Again, we have two different ways of deleting edges. The first possibility is
        \begin{equation}
    \begin{gathered}
    \begin{tikzpicture}[baseline={([yshift=2ex]current bounding box.center)},scale=0.8]
    \coordinate (a) at (0,0);
    \coordinate (b) at (1.5,0);
    \coordinate (c) at (0.75,0);
    \draw[thick] (c) circle (0.75);
    \draw[thick,fill=black] (a) circle (2pt)  (b) circle (2pt);
    \draw[Maroon, thick] (c) ellipse (1cm and 1cm);
    \node[scale=0.8] at (0.75,-1.5) {$\frac{1}{x_1+x_2}$};
    \draw[->,thick] (2.5,0) --++ (1,0);
    \begin{scope}[shift={(5,0)}]
        \coordinate (a) at (-0.75,0);
        \coordinate (b) at (0.75,0);
        \coordinate (c) at (0,0);
        \draw[thick] (c) circle (0.75);
        \draw[thick,fill=black] (a) circle (2pt)  (b) circle (2pt);
        \draw[Maroon, thick] (c) ellipse (1.1cm and 1.1cm);
        \draw[Maroon, thick] (0.95,0) arc (0:-180:0.95) arc (180:0:0.2) arc (0:180:-0.55) arc (0:-180:-0.2);
        \node[scale=0.8] at (0,-1.5) {$\frac{1}{x_1+x_2} \frac{1}{x_1+x_2+2 y_{12}}$};
        \draw[->,thick] (2.5,0) --++ (1,0);
    \end{scope}
    \begin{scope}[shift={(10,0)}]
        \coordinate (a) at (-0.75,0);
        \coordinate (b) at (0.75,0);
        \coordinate (c) at (0,0);
        \draw[thick] (c) circle (0.75);
        \draw[thick,fill=black] (a) circle (2pt)  (b) circle (2pt);
        \draw[Maroon, thick] (c) ellipse (1.1cm and 1.1cm);
        \draw[Maroon, thick] (0.95,0) arc (0:-180:0.95) arc (180:0:0.2) arc (0:180:-0.55) arc (0:-180:-0.2);
        \node[scale=0.8] at (0,-1.75) {$\substack{\frac{1}{x_1+x_2} \frac{1}{x_1+x_2+2 y_{12}}
        \\[3pt] 
        \hspace{1.2cm} \times
        \frac{1}{x_1+y_{12}+y_{21}} \frac{1}{x_2+y_{21}+y_{12}}}$};
        \draw[Maroon, thick] (a) ellipse (0.12cm and 0.12cm);
        \draw[Maroon, thick] (b) ellipse (0.12cm and 0.12cm);
    \end{scope}
    \end{tikzpicture}
    \end{gathered}
    \end{equation}
    and the second one is
           \begin{equation}
    \begin{gathered}
    \begin{tikzpicture}[baseline={([yshift=2ex]current bounding box.center)},scale=0.8]
    \coordinate (a) at (0,0);
    \coordinate (b) at (1.5,0);
    \coordinate (c) at (0.75,0);
    \draw[thick] (c) circle (0.75);
    \draw[thick,fill=black] (a) circle (2pt)  (b) circle (2pt);
    \draw[Maroon, thick] (c) ellipse (1cm and 1cm);
    \node[scale=0.8] at (0.75,-1.5) {$\frac{1}{x_1+x_2}$};
    \draw[->,thick] (2.5,0) --++ (1,0);
    \begin{scope}[shift={(5,0)}]
        \coordinate (a) at (-0.75,0);
        \coordinate (b) at (0.75,0);
        \coordinate (c) at (0,0);
        \draw[thick] (c) circle (0.75);
        \draw[thick,fill=black] (a) circle (2pt)  (b) circle (2pt);
        \draw[Maroon, thick] (c) ellipse (1.1cm and 1.1cm);
        \draw[Maroon, thick, yscale=-1] (0.95,0) arc (0:-180:0.95) arc (180:0:0.2) arc (0:180:-0.55) arc (0:-180:-0.2);
        \node[scale=0.8] at (0,-1.5) {$\frac{1}{x_1+x_2} \frac{1}{x_1+x_2+2 y_{21}}$};
        \draw[->,thick] (2.5,0) --++ (1,0);
    \end{scope}
    \begin{scope}[shift={(10,0)}]
        \coordinate (a) at (-0.75,0);
        \coordinate (b) at (0.75,0);
        \coordinate (c) at (0,0);
        \draw[thick] (c) circle (0.75);
        \draw[thick,fill=black] (a) circle (2pt)  (b) circle (2pt);
        \draw[Maroon, thick] (c) ellipse (1.1cm and 1.1cm);
        \draw[Maroon, thick, yscale=-1] (0.95,0) arc (0:-180:0.95) arc (180:0:0.2) arc (0:180:-0.55) arc (0:-180:-0.2);
        \node[scale=0.8] at (0,-1.75) {$\substack{\frac{1}{x_1+x_2} \frac{1}{x_1+x_2+2 y_{21}}
        \\[3pt] 
        \hspace{1.2cm} \times
        \frac{1}{x_1+y_{12}+y_{21}} \frac{1}{x_2+y_{21}+y_{12}}}$};
        \draw[Maroon, thick] (a) ellipse (0.12cm and 0.12cm);
        \draw[Maroon, thick] (b) ellipse (0.12cm and 0.12cm);
    \end{scope}
    \end{tikzpicture}
    \end{gathered}
    \end{equation}
The final result for the one-loop graph is therefore the sum of these two contributions:
    \begin{align}
    \begin{gathered}
    \begin{tikzpicture}[baseline={([yshift=2ex]current bounding box.center)},scale=0.8]
    \coordinate (a) at (0,0);
    \coordinate (b) at (1.5,0);
    \draw[thick] (0.75,0) circle (0.75);
    \draw[thick,fill=black] (a) circle (2pt) node[left] {$x_1$};
    \draw[thick,fill=black] (b) circle (2pt) node[right] {$x_2$};
    \node[] at (0.75,1) {$y_{12}$};
    \node[] at (0.75,-1) {$y_{21}$};
    \end{tikzpicture}
    \end{gathered}
    & =
    \frac{1}{(x_1+x_2)(x_1+y_{12}+y_{21}) (x_2+y_{12}+y_{21})}
    \nonumber
    \\ & \hspace{2cm} \times \left[ 
    \frac{1}{x_1+x_2+2 y_{12}} + 
    \frac{1}{x_2+x_2+2 y_{21}}
    \right] \,.
    \end{align}
\end{mdexample}

\section[Wavefunction from combinatorics]{Wavefunction from combinatorics\\
\normalfont{\textit{Paolo Benincasa}}}

\subsection{From graphs to polytopes}
One of the main messages from Sec.~\ref{sec:generalities_pert} is the one-to-one correspondence between subgraphs of a given graph $\mathcal{G}$ and singularities of $\psi_{\mathcal{G}}$. In addition, the existence of a recursion relation based on iteratively constructing subgraphs is combinatorial in nature. So, it becomes legitimate to ask whether there is some mathematical structure which determines it. Let us begin with the simplest example, the two-site tree graph, and let us consider the subgraph (and hence the singularities) associated to it,
\begin{equation}
    \begin{gathered}
    \begin{tikzpicture}[baseline={([yshift=2ex]current bounding box.center)}]
        \coordinate (a) at (0,0);
        \coordinate (b) at (1.5,0);
        \draw[thick] (a) -- (b);
        \draw[thick,fill=black] (a) circle (2pt) node[below,yshift=-4] {$x_1$}  (b) circle (2pt) node[below,yshift=-4] {$x_2$};
        \node[] at (0.75,0.35) {$y_{12}$};
        \draw[->,thick] (2.5,0) --++ (1,0);
    \begin{scope}[shift={(4.5,0.8)}]
        \coordinate (a) at (0,0);
        \coordinate (b) at (1.5,0);
        \draw[thick] (a) -- (b);
        \draw[thick,fill=black] (a) circle (2pt)(b) circle (2pt);
        \node[scale=0.8] at (3,0) {$x_1+x_2=0$};
        \draw[Maroon, thick] (0.75, 0) ellipse (0.95cm and 0.2cm);
    \end{scope}
    \begin{scope}[shift={(4.5,0)}]
        \coordinate (a) at (0,0);
        \coordinate (b) at (1.5,0);
        \draw[thick] (a) -- (b);
        \draw[thick,fill=black] (a) circle (2pt)(b) circle (2pt);
        \node[scale=0.8] at (3,0) {$x_1+y_{12}=0$};
        \draw[Maroon, thick] (a) ellipse (0.15cm and 0.15cm);
    \end{scope}
    \begin{scope}[shift={(4.5,-0.8)}]
        \coordinate (a) at (0,0);
        \coordinate (b) at (1.5,0);
        \draw[thick] (a) -- (b);
        \draw[thick,fill=black] (a) circle (2pt)(b) circle (2pt);
        \node[scale=0.8] at (3,0) {$y_{12}+x_2=0$};
        \draw[Maroon, thick] (b) ellipse (0.15cm and 0.15cm);
    \end{scope}
    \end{tikzpicture}
    \end{gathered}
\end{equation}
Note that the loci of the singularities are all homogeneous linear equations in $(x_1,y_{12},x_2)$. This implies that taking this triple as a local coordinate system in $\mathbb{P}^2$ , these three equations determine three lines. Choosing an orientation, the positive half-planes that they identify, i.e., taking $x_1+x_2 \geq 0$, $x_1+y_{12} \geq 0$, $y_{12} + x_2 \geq 0$ -- overlap on a triangle (the shaded area):
\begin{equation}
    \begin{gathered}
    \begin{tikzpicture}[baseline={([yshift=2ex,xshift=10ex]current bounding box.center)}]
        \coordinate (a) at (-0.5,0);
        \coordinate (b) at (0.5,0);
        \coordinate (c) at (0,0.866025);
        \draw[thick,fill=black!30] (a) node[below left,scale=0.8,xshift=-10] {$2$} -- (b) node[below right,scale=0.8,xshift=10] {$3$}   -- (c) node[above,scale=0.8,yshift=10] {$1$}  -- cycle;
        \draw[-<,thick] (a) --++ (60:0.5);
        \draw[->,thick] (a) --++ (0:0.5);
        \draw[-<,thick] (c) --++ (120:-0.5);
        \draw[thick] (a) --++ (60:1.5) (a) --++ (60:-0.5)
        (0,0) --++ (0:1) (0,0) --++ (180:1)
        (c) --++ (120:0.5) (c) --++ (120:-1.5);
        \node[scale=0.8] at ($(b)+(1.5,0)$) {$x_1+x_2=0$};
        \node[scale=0.8] at ($(a)+(-1,-0.8)$) {$y_{12}+x_2=0$};
        \node[scale=0.8] at ($(b)+(1,-0.8)$) {$x_1+y_{12}=0$};
    \end{tikzpicture}
    \end{gathered}
\end{equation}
Interestingly given a triangle $\mathcal{P}_{\mathcal{G}}$ whose sides are identified by $\mathcal{P}_{\mathcal{G}} \cap \mathcal{W}^{(\mathcal{G})}$ -- $\mathcal{W}^{(\mathcal{G})}$ being the normal covector to the line containing one of its sides -- there exists a unique (up to normalization) differential form associated to it such that it has only logarithmic singularities and such logarithmic singularities are associated to its boundaries (i.e.,  sides and vertices):
\begin{equation}
    \omega(\mathcal{Y}, \mathcal{P}_\mathcal{G}) = \frac{\langle 123 \rangle^2 \langle \mathcal{Y} \d^2 \mathcal{Y} \rangle}{\langle \mathcal{Y} 12 \rangle \langle \mathcal{Y} 23 \rangle \langle \mathcal{Y} 31 \rangle}\,,
\end{equation}
where $\mathcal{Y} : = (x_1, y_{12}, x_2) \in \mathbb{P}^2$ is a generic point in $\mathbb{P}^2$, $\langle \mathcal{Y} \d^2 \mathcal{Y} \rangle$ is the canonical measure in $\mathbb{P}^2$, $\langle y i (i+1) \rangle \equiv \epsilon_{I J K} y^{I} Z_{(i)}^J Z_{(i+1)}^{K}$, $\{ Z_{(i)}^I, \quad i = 1,2,3 \}$ being the set of vertices of the triangle and $\mathcal{W_I} = \epsilon_{IJK} Z_{(i)}^J Z_{(i+1)}^K$ being the covector that identifies the line passing through the two vertices $Z_{(i)}$ and $Z_{(i+1)}$. The numerator is, in this simple case, just determined by projectivity: the canonical form has to be invariant under $\GL(1)$ rescaling $\mathcal{Y} \to \lambda \mathcal{Y}$, $\{Z_{(i)} \to \lambda Z_{(i)}\}, \quad i=1,2,3$, $\lambda \in \mathbb{R}_+$. The coefficient of the canonical form -- named canonical function -- is nothing but the wavefunction coefficient associated to the two-site tree graph:
\begin{equation}
    \omega (\mathcal{Y}, \mathcal{P}_{\mathcal{G}}) = \Omega(\mathcal{Y}, \mathcal{P}_{\mathcal{G}}) \langle \mathcal{Y} \d^2 \mathcal{Y} \rangle \,,
\end{equation}
with
\begin{equation}
    \Omega(\mathcal{Y}, \mathcal{P}_{\mathcal{G}}) = \psi_{\mathcal{G}} (x,y) \,.
\end{equation}
The general statement is that given a (reduced) weighted graph $\mathcal{G}$, it is in one-to-one correspondence with a \emph{cosmological polytope}, whose canonical function encodes the wavefunction coefficient associated to $\mathcal{G}$. Going back to the two-site example, we can naturally associate a triangle to it by
\begin{itemize}
    \item[(i)] considering the local coordinates $\mathcal{Y} \equiv (x_1, y_{12},x_2) \in \mathbb{P}^2$ constructed out of its weights;
    \item[(ii)] in this system of local coordinate, considering the canonical basis for $\mathbb{R}^3$, $\vx_1\equiv(1,0,0)$, $\vec{y}_{12} \equiv (0,1,0)$, $\vx_{2} \equiv (0,0,1)$.
\end{itemize}
Then, the triple $(\vx_1,\vec{y}_{12},\vx_2)$ defines a triangle whose sides are characterized by $(\vx_1,\vec{y}_{12},\vx_2)$ as midpoints. The vertices of such a triangle are then given by
\begin{equation}
    \begin{gathered}
    \begin{tikzpicture}[baseline={([yshift=2ex,xshift=10ex]current bounding box.center)}]
        \coordinate (a) at (-0.5,0);
        \coordinate (b) at (0.5,0);
        \coordinate (c) at (0,0.866025);
        \draw[thick] (a) node[below left,scale=0.8,xshift=-5] {$Z_2$} -- (b) node[below right,scale=0.8,xshift=5] {$Z_3$}   -- (c) node[above,scale=0.8,yshift=5] {$Z_1$}  -- cycle;
        \draw[fill=black] (a) circle (2pt) (b) circle (2pt) (c) circle (2pt);
        \node[scale=0.8] at (-0.5,0.5) {$\bar{x}_1$};
        \node[scale=0.8] at (0.5,0.5) {$\bar{x}_2$};
        \node[scale=0.8] at (0,-0.3) {$\bar{y}_{12}$};
    \end{tikzpicture}
    \end{gathered}
    \hspace{2cm}
    \begin{aligned}
        Z_1 & = \vx_1 - \vec{y}_{12} + \vx_2 \,, \\
        Z_2 & = \vx_1 + \vec{y}_{12} - \vx_2 \,, \\
        Z_3 & = - \vx_1 + \vec{y}_{12} + \vx_2 \,,
    \end{aligned}
\end{equation}
and the associated canonical form is given by
\begin{equation}
\begin{split}
        \omega(\mathcal{Y}, \mathcal{P}_\mathcal{G})& = \frac{\langle 123 \rangle^2 \langle \mathcal{Y} \d^2 \mathcal{Y} \rangle}{\langle \mathcal{Y}12 \rangle \langle \mathcal{Y}23 \rangle \langle \mathcal{Y}31 \rangle}
        \\&
        =
        \underbracket[0.4pt]{{}\frac{1}{(x_1+x_2)(x_1+y_{12}) (y_{12} + x_2)} }_{\psi_{\mathcal{G}}(x,y)}
    \frac{\d x_1 \wedge \d y_{12} \wedge \d x_2}{\text{Vol} \{ \GL (1) \}} \,.
\end{split}
\end{equation}

\subsection{Cosmological Polytopes}
It is useful -- and perhaps more intuitive -- to keep the graphs as our starting point. A first observation is that any weighted graph can be obtained from a collection of two-site graphs by identifying some sites. For example, let us consider a collection of two two-site tree graphs:
\begin{equation}
    \begin{gathered}
    \begin{tikzpicture}[baseline={([yshift=2ex]current bounding box.center)}]
    \coordinate (a) at (0,0);
    \coordinate (b) at (1.5,0);
    \draw[thick] (a) -- (b);
    \draw[thick,fill=black] (a) circle (2pt) node[below] {$x_1$};
    \draw[thick,fill=black] (b) circle (2pt) node[below] {$x_2$};
    \node[] at (0.75,0.35) {$y_{12}$};
    \end{tikzpicture}
    \hspace{2cm}
    \begin{tikzpicture}[baseline={([yshift=2ex]current bounding box.center)}]
    \coordinate (a) at (0,0);
    \coordinate (b) at (1.5,0);
    \draw[thick] (a) -- (b);
    \draw[thick,fill=black] (a) circle (2pt) node[below] {$x_1'$};
    \draw[thick,fill=black] (b) circle (2pt) node[below] {$x_2'$};
    \node[] at (0.75,0.35) {$y_{12}'$};
    \end{tikzpicture}
    \end{gathered}
\end{equation}
It is possible to obtain from them new graphs from all the inequivalent ways of identifying one or more sites:
\begin{align}
    \text{1-site identification: } \quad  & 
    \begin{gathered}
    \begin{tikzpicture}[baseline={([yshift=2ex]current bounding box.center)}]
    \coordinate (a) at (0,0);
    \coordinate (b) at (1.5,0);
    \coordinate (c) at (3,0);
    \draw[thick] (a) -- (b);
    \draw[thick] (b) -- (c);
    \draw[thick,fill=black] (a) circle (2pt) node[below] {$x_1{\color{white} '}$};
    \draw[thick,fill=black] (b) circle (2pt) node[below] {$x_2=x_2'$};
    \draw[thick,fill=black] (c) circle (2pt) node[below] {$x_1'$};
    \node[] at (0.75,0.35) {$y_{12}$};
    \node[] at (2.25,0.35) {$y_{12}'$};
    \end{tikzpicture}
    \end{gathered}
    \\ 
    \text{2-site identification: } \quad  &
    \begin{gathered}
    \begin{tikzpicture}[baseline={([yshift=2ex]current bounding box.center)}]
    \coordinate (a) at (0,0);
    \coordinate (b) at (1.5,0);
    \draw[thick] (0.75,0) circle (0.75);
    \draw[thick,fill=black] (a) circle (2pt) node[left] {$x_1{=}x_1'$};
    \draw[thick,fill=black] (b) circle (2pt) node[right] {$x_2{=}x_2'$};
    \node[] at (0.75,1) {$y_{12}$};
    \node[] at (0.75,-1) {$y_{12}'$};
    \end{tikzpicture}
    \end{gathered}
\end{align}
As in each 2-site tree graph there is a triangle associated to it; given two 2-site graphs, it is possible to embed the associated triangles in the same space $\mathbb{P}^5$, where they are still disconnected. The identification of a pair of sites belonging to different 2-site tree graphs, say $x_2 = x_2'$, corresponds geometrically to the identification of the vertices $\vx_2 = \vx_2'$. In terms of the vertices of the triangles, this implies a linear relation among two pairs of them, i.e., the pair having as a midpoint $\vx_2$ and the one having as a midpoint $\vx_2'$,
\begin{equation}
    Z_3+Z_1 \sim Z_3' + Z_1' \,.
\end{equation}
Such a relation identifies a 2-plane -- there are as many of such 2-planes as shared sites in a graph. If the 2-site graphs have the following triple associated to them,
\begin{align}
\begin{gathered}
    \begin{tikzpicture}[baseline={([yshift=2ex]current bounding box.center)}]
    \coordinate (a) at (0,0);
    \coordinate (b) at (1.5,0);
    \draw[thick] (a) -- (b);
    \draw[thick,fill=black] (a) circle (2pt) node[below] {$x_1$};
    \draw[thick,fill=black] (b) circle (2pt) node[below] {$x_2$};
    \node[] at (0.75,0.35) {$y_{12}$};
    \end{tikzpicture}
\end{gathered}
    &\hspace{.5cm}
    \{ \bf{x}_1 - \bf{y}_{12} + \bf{x}_2 ,
    \bf{x}_1 + \bf{y}_{12} - \bf{x}_2 ,
    -\bf{x}_1 + \bf{y}_{12} + \bf{x}_2 \}\,,
    \\
    \begin{gathered}
    \begin{tikzpicture}[baseline={([yshift=2ex]current bounding box.center)}]
    \coordinate (a) at (0,0);
    \coordinate (b) at (1.5,0);
    \draw[thick] (a) -- (b);
    \draw[thick,fill=black] (a) circle (2pt) node[below] {$x_1'$};
    \draw[thick,fill=black] (b) circle (2pt) node[below] {$x_2'$};
    \node[] at (0.75,0.35) {$y_{12}'$};
    \end{tikzpicture}
    \end{gathered}
    &\hspace{.5cm}
    \{ \bf{x}_1' - \bf{y}_{12}' + \bf{x}_2' ,
    \bf{x}_1' + \bf{y}_{12}' - \bf{x}_2' ,
    -\bf{x}_1' + \bf{y}_{12}'+ \bf{x}_2' \}\,,
\end{align}
then the two graph topologies that can be constructed out of these 2-site graphs, have each associated a cosmological polytope with 6 vertices given by
\begin{align}
    \begin{gathered}
    \begin{tikzpicture}[baseline={([yshift=2ex]current bounding box.center)}]
    \coordinate (a) at (0,0);
    \coordinate (b) at (1.5,0);
    \coordinate (c) at (3,0);
    \draw[thick] (a) -- (b);
    \draw[thick] (b) -- (c);
    \draw[thick,fill=black] (a) circle (2pt) node[below] {$x_1{\color{white} '}$};
    \draw[thick,fill=black] (b) circle (2pt) node[below] {$x_2$};
    \draw[thick,fill=black] (c) circle (2pt) node[below] {$x_1'$};
    \node[] at (0.75,0.35) {$y_{12}$};
    \node[] at (2.25,0.35) {$y_{12}'$};
    \node[yshift=-30] at (b) {$x_2 = x_2'$};
    \end{tikzpicture}
    \end{gathered}
    \quad
    & \{ \bf{x}_1 - \bf{y}_{12} + \bf{x}_2 ,
    \bf{x}_1 + \bf{y}_{12} - \bf{x}_2 ,
    -\bf{x}_1 + \bf{y}_{12} + \bf{x}_2,
    \\
    & \hspace{0.5cm} \bf{x}_1' - \bf{y}_{12}' + \bf{x}_2 ,
    \bar{x}_1' + \bar{y}_{12}' - \bf{x}_2 ,
    -\bf{x}_1' + \bf{y}_{12}'+ \bf{x}_2 \} \subset \mathbb{P}^4\,,
    \nonumber
    \\
    \begin{gathered}
    \begin{tikzpicture}[baseline={([yshift=2ex]current bounding box.center)}]
    \coordinate (a) at (0,0);
    \coordinate (b) at (1.5,0);
    \draw[thick] (0.75,0) circle (0.75);
    \draw[thick,fill=black] (a) circle (2pt) node[left] {$x_1$};
    \draw[thick,fill=black] (b) circle (2pt) node[right] {$x_2$};
    \node[] at (0.75,1) {$y_{12}$};
    \node[] at (0.75,-1) {$y_{12}'$};
    \end{tikzpicture}
    \end{gathered}
    \quad
    & \{ \bf{x}_1 - \bf{y}_{12} + \bf{x}_2 ,
    \bf{x}_1 + \bf{y}_{12} - \bf{x}_2 ,
    -\bf{x}_1 + \bf{y}_{12} + \bf{x}_2,
    \\
    & \hspace{0.5cm} \bf{x}_1 - \bf{y}_{12}' + \bf{x}_2 ,
    \bf{x}_1 + \bf{y}_{12}' - \bf{x}_2 ,
    -\bf{x}_1 + \bf{y}_{12}'+ \bf{x}_2 \} \subset \mathbb{P}^3\,,
    \nonumber
\end{align}
More generally, given an ordinary, connected, graph $\mathcal{G}$ with $n_s$ sites and $n_e$ edges, the associated cosmological polytope lives in $\mathbb{P}^{n_s+n_e-1}$ with local coordinates given by all the weights of the graph $\mathcal{Y}: = ( \{x_s\}_{s \in \mathcal{V}}, \{ y_e\}_{e \in \mathcal{E}}$, and is the convex hull of $3^{n_e}$ vertices:
\begin{equation}
    \{ \mathbf{x}_{s_e} - \mathbf{y}_e + \mathbf{x}_{s_e'} \}_{e \in \mathcal{E}} \,.
\end{equation}
The correspondence between graphs and cosmological polytopes allows keeping track of the vertex structure of all its boundaries. This can be obtained by introducing a marking that identifies those vertices which \emph{are not} on the facet -- a facet is the codimension-1 face of the cosmological polyotope identified by an equation which corresponds to one of the poles of the wavefunction and which are associated to a subgraph.

The marking is given by
\begin{equation}
\begin{gathered}
    \begin{tikzpicture}[baseline={([yshift=2ex]current bounding box.center)},cross/.style={cross out, draw, minimum size=2*(#1-\pgflinewidth), inner sep=0pt, outer sep=0pt}]
    \coordinate (a) at (0,0);
    \coordinate (b) at (1.5,0);
    \draw[thick] (a) -- (b);
    \draw[thick,fill=black] (a) circle (2pt) node[below] {$x_{s_e}$};
    \draw[thick,fill=black] (b) circle (2pt) node[below] {$x_{s'_e}$};
    \node[] at (0.75,0.35) {$y_{e}$};
    \node[very thick, cross=4pt, rotate=0, color=RoyalBlue, scale=.625] at (0.75,0) {};
    \node[scale=0.8] at (0.75,-1) {$\mathcal{W} \cdot \left( \mathbf{x}_{s_e} {-} \mathbf{y}_e {+} \mathbf{x}_{s'_e} \right) > 0$};
    \end{tikzpicture}
    \hspace{1cm}
        \begin{tikzpicture}[baseline={([yshift=2ex]current bounding box.center)},cross/.style={cross out, draw, minimum size=2*(#1-\pgflinewidth), inner sep=0pt, outer sep=0pt}]
    \coordinate (a) at (0,0);
    \coordinate (b) at (1.5,0);
    \draw[thick] (a) -- (b);
    \draw[thick,fill=black] (a) circle (2pt) node[below] {$x_{s_e}$};
    \draw[thick,fill=black] (b) circle (2pt) node[below] {$x_{s'_e}$};
    \node[] at (0.75,0.35) {$y_{e}$};
    \node[very thick, cross=4pt, rotate=0, color=RoyalBlue, scale=.625] at (0.2,0) {};
    \node[scale=0.8] at (0.75,-1) {$\mathcal{W} \cdot \left( \mathbf{x}_{s_e} {+} \mathbf{y}_e {-} \mathbf{x}_{s'_e} \right) > 0$};
    \end{tikzpicture}
    \hspace{1cm}
        \begin{tikzpicture}[baseline={([yshift=2ex]current bounding box.center)},cross/.style={cross out, draw, minimum size=2*(#1-\pgflinewidth), inner sep=0pt, outer sep=0pt}]
    \coordinate (a) at (0,0);
    \coordinate (b) at (1.5,0);
    \draw[thick] (a) -- (b);
    \draw[thick,fill=black] (a) circle (2pt) node[below] {$x_{s_e}$};
    \draw[thick,fill=black] (b) circle (2pt) node[below] {$x_{s'_e}$};
    \node[] at (0.75,0.35) {$y_{e}$};
    \node[very thick, cross=4pt, rotate=0, color=RoyalBlue, scale=.625] at (1.3,0) {};
    \node[scale=0.8] at (0.75,-1) {$\mathcal{W} \cdot \left( {-} \mathbf{x}_{s_e} {+} \mathbf{y}_e {+} \mathbf{x}_{s'_e} \right) > 0$};
    \end{tikzpicture}
\end{gathered}
\end{equation}

A vertex $Z$ is on a facet identified by a covector $\mathcal{W}$ if $Z^I \mathcal{W}_I = 0$. If it is not on the facet, then $Z^I \mathcal{W}_I > 0$. This implies that to know which vertex is on a facet, one would have to check whether $Z^I \mathcal{W}_I = 0$. A graphical rule in the marking allows us to find the answer immediately. 

The facet of $\mathcal{P}_{\mathcal{G}}$ identified by the subgraph $\mathfrak{g} \subseteq \mathcal{G}$ is given by marking all the internal edges in the middle, and all the edges that depart from this subgraph on the side close to $\mathfrak{g}$:
\begin{equation*}
    \begin{tikzpicture}[ball/.style = {circle, draw, align=center, anchor=north, inner sep=0}, cross/.style={cross out, draw, minimum size=2*(#1-\pgflinewidth), inner sep=0pt, outer sep=0pt}, scale=1.25, transform shape]
  \begin{scope}
\coordinate (v1) at (0,0);
\node[below, scale=0.5] at ([yshift=-1mm]v1) {$x_1$};
\coordinate (v2) at ($(v1)+(0,1.25)$);
\node[above, scale=0.5] at ([yshift=1mm]v2) {$x_2$};
\coordinate (v3) at ($(v2)+(1.25,0)$);
\node[above, scale=0.5] at ([yshift=1mm]v3) {$x_3$};
\coordinate (v4) at ($(v3)+(1.25,0)$);
\node[above, scale=0.5] at ([yshift=1mm]v4) {$x_4$};
\coordinate (v5) at ($(v4)-(0,0.625)$);
\node[right, scale=0.5] at ([xshift=1mm]v5) {$x_5$};
\coordinate (v6) at ($(v5)-(0,0.625)$);
\node[below, scale=0.5] at ([yshift=-1mm]v6) {$x_6$};
\coordinate (v7) at ($(v6)-(1.25,0)$);
\node[below, scale=0.5] at ([yshift=-1mm]v7) {$x_7$};
   \draw[thick] (v1) -- (v2) -- (v3) -- (v4) -- (v5) -- (v6) -- (v7) -- cycle;
   \draw[thick] (v3) -- (v7);
   \draw[fill=black] (v1) circle (2pt);
   \draw[fill=black] (v2) circle (2pt);
   \draw[fill=black] (v3) circle (2pt);
   \draw[fill=black] (v4) circle (2pt);
   \draw[fill=black] (v5) circle (2pt);
   \draw[fill=black] (v6) circle (2pt);
   \draw[fill=black] (v7) circle (2pt);
   \coordinate (v12) at ($(v1)!0.5!(v2)$);
   \coordinate (v23) at ($(v2)!0.5!(v3)$);
   \coordinate (v34) at ($(v3)!0.5!(v4)$);
   \coordinate (v45) at ($(v4)!0.5!(v5)$);
   \coordinate (v56) at ($(v5)!0.5!(v6)$);
   \coordinate (v67) at ($(v6)!0.5!(v7)$);
   \coordinate (v71) at ($(v7)!0.5!(v1)$);
   \coordinate (v37) at ($(v3)!0.5!(v7)$);
   \node[very thick, cross=4pt, rotate=0, color=RoyalBlue, scale=.625] at (v12) {};
   \node[very thick, cross=4pt, rotate=0, color=RoyalBlue, scale=.625] at (v23) {};
   \node[very thick, cross=4pt, rotate=0, color=RoyalBlue, scale=.625] at (v34) {};
   \node[very thick, cross=4pt, rotate=0, color=RoyalBlue, scale=.625] at (v45) {};
   \node[very thick, cross=4pt, rotate=0, color=RoyalBlue, scale=.625] at (v56) {};
   \node[very thick, cross=4pt, rotate=0, color=RoyalBlue, scale=.625] at (v67) {};
   \node[very thick, cross=4pt, rotate=0, color=RoyalBlue, scale=.625] at (v71) {};
   \node[very thick, cross=4pt, rotate=0, color=RoyalBlue, scale=.625] at (v37) {};
   			\coordinate (n2w) at ($(v2)-(.125,0)$);
			\coordinate (n2n) at ($(v2)+(0,.125)$);
			\coordinate (n3n) at ($(v3)+(0,.175)$);
			\coordinate (n4n) at ($(v4)+(0,.125)$);
			\coordinate (n4e) at ($(v4)+(.125,0)$);
			\coordinate (n5e) at ($(v5)+(.175,0)$);
			\coordinate (n6e) at ($(v6)+(.125,0)$);
			\coordinate (n6s) at ($(v6)-(0,.125)$);
			\coordinate (n7s) at ($(v7)-(0,.175)$);
			\coordinate (n1s) at ($(v1)-(0,.125)$);
			\coordinate (n1w) at ($(v1)-(.125,0)$);
			\coordinate (ntw) at ($(v1)!0.5!(v2)-(.175,0)$);
			\draw[thick, RoyalBlue] plot [smooth cycle] coordinates {(n2w) (n2n) (n3n) (n4n) (n4e) (n5e) (n6e) (n6s) (n7s) (n1s) (n1w) (ntw)};
      \node[below=.18cm of v7, scale=0.8, color=RoyalBlue] {\footnotesize $\displaystyle\mathfrak{g} = \mathcal{G}$};
  \end{scope}
  \begin{scope}[shift={(3.5,0)}, transform shape]
\coordinate (v1) at (0,0);
\node[below, scale=0.5] at ([yshift=-1mm]v1) {$x_1$};
\coordinate (v2) at ($(v1)+(0,1.25)$);
\node[above, scale=0.5] at ([yshift=1mm]v2) {$x_2$};
\coordinate (v3) at ($(v2)+(1.25,0)$);
\node[above, scale=0.5] at ([yshift=1mm]v3) {$x_3$};
\coordinate (v4) at ($(v3)+(1.25,0)$);
\node[above, scale=0.5] at ([yshift=1mm]v4) {$x_4$};
\coordinate (v5) at ($(v4)-(0,0.625)$);
\node[right, scale=0.5] at ([xshift=1mm]v5) {$x_5$};
\coordinate (v6) at ($(v5)-(0,0.625)$);
\node[below, scale=0.5] at ([yshift=-1mm]v6) {$x_6$};
\coordinate (v7) at ($(v6)-(1.25,0)$);
\node[below, scale=0.5] at ([yshift=-1mm]v7) {$x_7$};
   \draw[thick] (v1) -- (v2) -- (v3) -- (v4) -- (v5) -- (v6) -- (v7) -- cycle;
   \draw[thick] (v3) -- (v7);
   \draw[fill=black] (v1) circle (2pt);
   \draw[fill=black] (v2) circle (2pt);
   \draw[fill=black] (v3) circle (2pt);
   \draw[fill=black] (v4) circle (2pt);
   \draw[fill=black] (v5) circle (2pt);
   \draw[fill=black] (v6) circle (2pt);
   \draw[fill=black] (v7) circle (2pt);
   \coordinate (v12) at ($(v1)!0.5!(v2)$);
   \coordinate (v23) at ($(v2)!0.5!(v3)$);
   \coordinate (v34) at ($(v3)!0.5!(v4)$);
   \coordinate (v45) at ($(v4)!0.5!(v5)$);
   \coordinate (v56) at ($(v5)!0.5!(v6)$);
   \coordinate (v67) at ($(v6)!0.5!(v7)$);
   \coordinate (v71) at ($(v7)!0.5!(v1)$);
   \coordinate (v37) at ($(v3)!0.5!(v7)$);
   \node[very thick, cross=4pt, rotate=0, color=Maroon, scale=.625] at (v34) {};
   \node[very thick, cross=4pt, rotate=0, color=Maroon, scale=.625] at (v45) {};
   \node[very thick, cross=4pt, rotate=0, color=Maroon, scale=.625, left=.15cm of v3] (v3l) {};
   \node[very thick, cross=4pt, rotate=0, color=Maroon, scale=.625, below=.15cm of v3] (v3b) {};
   \node[very thick, cross=4pt, rotate=0, color=Maroon, scale=.625, below=.1cm of v5] (v5b){};
   \coordinate (a) at ($(v3l)!0.5!(v3)$);
   \coordinate (b) at ($(v3)+(0,.125)$);
   \coordinate (c) at ($(v34)+(0,.175)$);
   \coordinate (d) at ($(v4)+(0,.125)$);
   \coordinate (e) at ($(v4)+(.125,0)$);
   \coordinate (f) at ($(v45)+(.175,0)$);
   \coordinate (g) at ($(v5)+(.125,0)$);
   \coordinate (h) at ($(v5b)!0.5!(v5)$);
   \coordinate (i) at ($(v5)-(.125,0)$);
   \coordinate (j) at ($(v45)-(.175,0)$);
   \coordinate (k) at ($(v34)-(0,.175)$);
   \coordinate (l) at ($(v3)-(0,.125)$);
   \draw [thick, Maroon] plot [smooth cycle] coordinates {(a) (b) (c) (d) (e) (f) (g) (h) (i) (j) (k) (l)};
   \node[below=.18cm of v7, scale=0.8, color=Maroon] {\footnotesize $\displaystyle\mathfrak{g} \subset \mathcal{G}$};
  \end{scope}
    \begin{scope}[shift={(7,0)}, transform shape]
\coordinate (v1) at (0,0);
\node[below, scale=0.5] at ([yshift=-1mm]v1) {$x_1$};
\coordinate (v2) at ($(v1)+(0,1.25)$);
\node[above, scale=0.5] at ([yshift=1mm]v2) {$x_2$};
\coordinate (v3) at ($(v2)+(1.25,0)$);
\node[above, scale=0.5] at ([yshift=1mm]v3) {$x_3$};
\coordinate (v4) at ($(v3)+(1.25,0)$);
\node[above, scale=0.5] at ([yshift=1mm]v4) {$x_4$};
\coordinate (v5) at ($(v4)-(0,0.625)$);
\node[right, scale=0.5] at ([xshift=1mm]v5) {$x_5$};
\coordinate (v6) at ($(v5)-(0,0.625)$);
\node[below, scale=0.5] at ([yshift=-1mm]v6) {$x_6$};
\coordinate (v7) at ($(v6)-(1.25,0)$);
\node[below, scale=0.5] at ([yshift=-1mm]v7) {$x_7$};
   \draw[thick] (v1) -- (v2) -- (v3) -- (v4) -- (v5) -- (v6) -- (v7) -- cycle;
   \draw[thick] (v3) -- (v7);
   \draw[fill=black] (v1) circle (2pt);
   \draw[fill=black] (v2) circle (2pt);
   \draw[fill=black] (v3) circle (2pt);
   \draw[fill=black] (v4) circle (2pt);
   \draw[fill=black] (v5) circle (2pt);
   \draw[fill=black] (v6) circle (2pt);
   \draw[fill=black] (v7) circle (2pt);
   \coordinate (v12) at ($(v1)!0.5!(v2)$);
   \coordinate (v23) at ($(v2)!0.5!(v3)$);
   \coordinate (v34) at ($(v3)!0.5!(v4)$);
   \coordinate (v45) at ($(v4)!0.5!(v5)$);
   \coordinate (v56) at ($(v5)!0.5!(v6)$);
   \coordinate (v67) at ($(v6)!0.5!(v7)$);
   \coordinate (v71) at ($(v7)!0.5!(v1)$);
   \coordinate (v37) at ($(v3)!0.5!(v7)$);
   \node[very thick, cross=4pt, rotate=0, color=Maroon, scale=.625] at (v37) {};
   \node[very thick, cross=4pt, rotate=0, color=Maroon, scale=.625, right=.15cm of v3] (v3r) {};
   \node[very thick, cross=4pt, rotate=0, color=Maroon, scale=.625, left=.15cm of v3] (v3l) {};
   \node[very thick, cross=4pt, rotate=0, color=Maroon, scale=.625, left=.15cm of v7] (v7l) {};
   \node[very thick, cross=4pt, rotate=0, color=Maroon, scale=.625, right=.15cm of v7] (v7r){};
   \draw [thick, Maroon] (v37) ellipse (.125cm and .75cm);
   \node[below=.18cm of v7, scale=0.8, color=Maroon] {\footnotesize $\displaystyle\mathfrak{g} \subset \mathcal{G}$};
  \end{scope}
 \end{tikzpicture}
\end{equation*}
Due to the one-to-one mapping between subgraphs and facets, the latter can be classified or listed using the former. The canonical functions of a cosmological polytope can generally be written as follows:
\begin{equation}
    \omega(\mathcal{Y}, \mathcal{P}_{\mathcal{G}}) = \frac{\mathfrak{n}_{\delta} (\mathcal{Y})}{\displaystyle\prod_{\mathfrak{g} \subseteq \mathcal{G}}
    q_\mathfrak{g}(\mathcal{Y})
    }
    \langle \mathcal{Y} \d^{n_s+n_e} \mathcal{Y} \rangle\,,
\end{equation}
where $q_\mathfrak{g} \equiv \mathcal{Y} \cdot \mathcal{W}^{(g)}$ and $\mathfrak{n}_{\delta}(\mathcal{Y})$ is a polynomial of degree $\delta$, with $\delta$ fixed by $\GL (1)$ invariance to be $\delta = \tilde{\nu}-n_s - n_e$ ($\tilde{\nu} \equiv \text{ number of facets}$). It turns out that also the polynomial $\mathfrak{n}_{\delta} (\mathcal{Y})$ has a geometrical interpretation: it is the locus of the intersections of the hyperplanes containing the facet of $\mathcal{P}_{\mathcal{G}}$ outside $\mathcal{P}_{\mathcal{G}}$ -- in the mathematical literature $\mathfrak{n}_{\delta} (\mathcal{Y}) = 0$ defines the adjoint surface.

A toy example can be given by a square in $\mathbb{P}^2$:
\begin{equation}
\begin{gathered}
    \begin{tikzpicture}[baseline={([yshift=2ex]current bounding box.center)},cross/.style={cross out, draw, minimum size=2*(#1-\pgflinewidth), inner sep=0pt, outer sep=0pt}]
    \coordinate (a) at (0,0);
    \coordinate (b) at (0.5,1.2);
    \coordinate (c) at (1.5,0.8);
    \coordinate (d) at (2.3,0);
    \draw[thick, fill=black!30] (a) node[left,scale=0.8] {$1$} -- (b) node[above left,scale=0.8] {$2$}  -- (c) node[above right,scale=0.8,xshift=-8pt] {$3$}  -- (d) node[right,scale=0.8] {$4$}  -- cycle;
    \draw[dashed] (b) --++ (67.3801:1);
    \draw[dashed] (d) --++ (0:2);
    \draw[dashed] (c) --++ (-21.8014:3);
    \draw[dashed] (c) --++ (-45:-1.7);
    \draw[fill=black] (0.6765,1.6235) circle (2pt) node[above,scale=0.8,yshift=3pt] {$A$};
    \draw[fill=black] (3.5,0.0) circle (2pt) node[below,scale=0.8] {$B$};
    \draw[Maroon,thick] (0.6765,1.6235) -- (3.5,0.0);
    \draw[Maroon,thick] ($(0.6765,1.6235)!-0.3!(3.5,0.0)$) -- ($(0.6765,1.6235)!1.3!(3.5,0.0)$);
    \draw[thick, ->] ($(a)!0.3!(b)$) -- ($(a)!0.6!(b)$);
    \draw[thick, ->] ($(b)!0.3!(c)$) -- ($(b)!0.6!(c)$);
    \draw[thick, ->] ($(c)!0.3!(d)$) -- ($(c)!0.6!(d)$);
    \draw[thick, ->] ($(d)!0.3!(a)$) -- ($(d)!0.6!(a)$);
    \end{tikzpicture}
\end{gathered}
\end{equation}
The square is the convex hull of the vertices $1,2,3,4$ and is characterized by the following canonical form
\begin{equation}
    \omega (\mathcal{Y},\mathcal{P}) =
    \frac{\mathfrak{n}_{\delta}(\mathcal{Y}) \langle \mathcal{Y} \d^2 \mathcal{Y} \rangle}{\langle \mathcal{Y}12 \rangle \langle \mathcal{Y}23 \rangle \langle \mathcal{Y}34 \rangle \langle \mathcal{Y}41 \rangle} \,.
\end{equation}
Because of $\GL (1)$ invariance, the numerator ought to be a polynomial of degree $\delta=1$. If higher codimension faces are defined as intersections of lower-codimension ones, -- and hence manifest themselves in terms of non-vanishing multiple residue of the canonical form -- there are further intersections which do not define a higher codimension face (like the points $A$ and $B$ for the quadrilateral above): the sequential residue of the canonical form has to vanish.
This implies that these intersections define subspaces of the adjoint. The numerator can be computed knowing these intersections. In the case of the quadrilateral above,
\begin{equation}
    Z_A^I = \epsilon^{IJK} \mathcal{W}_J^{(12)} \mathcal{W}_K^{(34)} \,, \qquad Z_B^I = \epsilon^{IJK} \mathcal{W}_J^{(23)} \mathcal{W}_K^{(41)}\,,
\end{equation}
where $\mathcal{W}_J^{(ij)}$ identifies the line through the vertices $Z_{(i)}^I$ and $Z_{(j)}^I$ and 
\begin{equation}
    \mathfrak{n}_{1} (\mathcal{Y}) = \mathcal{Y}^{I} \epsilon_{IJK} Z^J_A Z^K_B \,.
\end{equation}
The knowledge of these intersections determine also compatibility conditions for the facets. While for the simple example of the square is trivial, determining all the compatibility conditions of all codimension is not an easy task. In the case of the cosmological polytopes, the correspondence between graphs and polytopes, together with the markings, allow unveiling all this structure. From a physics perspective, these conditions imply which sequential discontinuities are non-zero. In codimension-two, they allowed to determine the analog for the wavefunction of the Steinmann relations \cite{Benincasa:2020aoj}. The higher-codimension one allows for novel ways of representing the wavefunction without introducing spurious poles \cite{Benincasa:2021qcb}. In order to fix the ideas, let us go back to the example of the quadrilateral. Rather than fixing the numerator, one could determine the canonical form from a triangulation,
\begin{equation}
    \omega (\mathcal{Y}, \mathcal{P})
    =
    \omega (\mathcal{Y}, \mathcal{P}_{124}) + 
    \omega (\mathcal{Y}, \mathcal{P}_{423})
    = 
    \omega (\mathcal{Y}, \mathcal{P}_{123}) + 
    \omega (\mathcal{Y}, \mathcal{P}_{341}) \,.
\end{equation}
\begin{equation}
\begin{gathered}
    \begin{tikzpicture}[baseline={([yshift=2ex]current bounding box.center)},cross/.style={cross out, draw, minimum size=2*(#1-\pgflinewidth), inner sep=0pt, outer sep=0pt}]
    \coordinate (a) at (0,0);
    \coordinate (b) at (0.5,1.2);
    \coordinate (c) at (1.5,1);
    \coordinate (d) at (2.3,0);
    \draw[thick] (a) node[left,scale=0.8] {$1$} -- (b) node[above left,scale=0.8] {$2$}  -- (c) node[above right,scale=0.8] {$3$}  -- (d) node[right,scale=0.8] {$4$}  -- cycle;
    \draw[thick] (b) -- (d); 
    \draw[thick, ->] ($(a)!0.3!(b)$) -- ($(a)!0.6!(b)$);
    \draw[thick, ->] ($(b)!0.3!(c)$) -- ($(b)!0.6!(c)$);
    \draw[thick, ->] ($(c)!0.3!(d)$) -- ($(c)!0.6!(d)$);
    \draw[thick, ->] ($(d)!0.3!(a)$) -- ($(d)!0.6!(a)$);
    \draw[thick, -<] ($(b)!0.3!(d)$) -- ($(b)!0.6!(d)$);
    \draw[thick, ->] ($(b)!0.3!(d)$) -- ($(b)!0.4!(d)$);
    \end{tikzpicture}
    \hspace{2cm}
        \begin{tikzpicture}[baseline={([yshift=2ex]current bounding box.center)},cross/.style={cross out, draw, minimum size=2*(#1-\pgflinewidth), inner sep=0pt, outer sep=0pt}]
    \coordinate (a) at (0,0);
    \coordinate (b) at (0.5,1.2);
    \coordinate (c) at (1.5,1);
    \coordinate (d) at (2.3,0);
    \draw[thick] (a) node[left,scale=0.8] {$1$} -- (b) node[above left,scale=0.8] {$2$}  -- (c) node[above right,scale=0.8] {$3$}  -- (d) node[right,scale=0.8] {$4$}  -- cycle;
    \draw[thick] (a) -- (c); 
    \draw[thick, ->] ($(a)!0.3!(b)$) -- ($(a)!0.6!(b)$);
    \draw[thick, ->] ($(b)!0.3!(c)$) -- ($(b)!0.6!(c)$);
    \draw[thick, ->] ($(c)!0.3!(d)$) -- ($(c)!0.6!(d)$);
    \draw[thick, ->] ($(d)!0.3!(a)$) -- ($(d)!0.6!(a)$);
    \draw[thick, -<] ($(a)!0.3!(c)$) -- ($(a)!0.7!(c)$);
    \draw[thick, ->] ($(a)!0.3!(c)$) -- ($(a)!0.5!(c)$);
    \end{tikzpicture}
\end{gathered}
\end{equation}
Note that in these triangulations, one makes use of a new codimension-1 boundary (the segments $24$ and $13$). 

From the perspective of the canonical forms as well as the codimension-1 boundary correspond to a singularity, the presence of $(24)$ and $(13)$ introduce a spurious singularity in each of the two representations.

In order to avoid the introduction of spurious boundaries/singularities, one should use only triangulations via the actual facets of the polytope of interest. In the case of the quadrilateral,
\begin{equation}
    \omega(\mathcal{Y},\mathcal{P})
    =
    \omega(\mathcal{Y},\mathcal{P}_{1A4}) + \omega(\mathcal{Y},\mathcal{P}_{3A2})
    =
     \omega(\mathcal{Y},\mathcal{P}_{12B}) + \omega(\mathcal{Y},\mathcal{P}_{B34}) \,,
\end{equation}
\begin{equation}
\begin{gathered}
    \begin{tikzpicture}[baseline={([yshift=2ex]current bounding box.center)},cross/.style={cross out, draw, minimum size=2*(#1-\pgflinewidth), inner sep=0pt, outer sep=0pt}]
    \coordinate (a) at (0,0);
    \coordinate (b) at (0.5,1.2);
    \coordinate (c) at (1.5,0.8);
    \coordinate (d) at (2.3,0);
    \draw[fill=black!30] (a) -- (d) -- (0.6765,1.6235) -- cycle;
    \draw[thick, fill=black!30] (a) node[below left,scale=0.8] {$1$} -- (b) node[above left,scale=0.8] {$2$}  -- (c) node[above right,scale=0.8,xshift=-8pt] {$3$}  -- (d) node[below right,scale=0.8] {$4$}  -- cycle;
    \draw[dashed] (b) --++ (67.3801:1);
    \draw[dashed] (d) --++ (0:2);
    \draw[dashed] (c) --++ (-21.8014:3);
    \draw[dashed] (c) --++ (-45:-1.7);
    \draw[fill=black] (0.6765,1.6235) circle (2pt) node[above,scale=0.8,yshift=3pt] {$A$};
    \draw[fill=black] (3.5,0.0) circle (2pt) node[below,scale=0.8] {$B$};
    \draw[Maroon,thick] (0.6765,1.6235) -- (3.5,0.0);
    \draw[Maroon,thick] ($(0.6765,1.6235)!-0.3!(3.5,0.0)$) -- ($(0.6765,1.6235)!1.3!(3.5,0.0)$);
    \draw[thick, ->] ($(a)!0.3!(b)$) -- ($(a)!0.6!(b)$);
    \draw[thick, ->] ($(b)!0.3!(c)$) -- ($(b)!0.6!(c)$);
    \draw[thick, ->] ($(c)!0.3!(d)$) -- ($(c)!0.6!(d)$);
    \draw[thick, ->] ($(d)!0.3!(a)$) -- ($(d)!0.6!(a)$);
    \end{tikzpicture}
    \hspace{1cm}
    \begin{tikzpicture}[baseline={([yshift=2ex]current bounding box.center)},cross/.style={cross out, draw, minimum size=2*(#1-\pgflinewidth), inner sep=0pt, outer sep=0pt}]
    \coordinate (a) at (0,0);
    \coordinate (b) at (0.5,1.2);
    \coordinate (c) at (1.5,0.8);
    \coordinate (d) at (2.3,0);
    \draw[fill=black!30] (c) -- (d) -- (3.5,0.0) -- cycle;
    \draw[thick, fill=black!30] (a) node[below left,scale=0.8] {$1$} -- (b) node[above left,scale=0.8] {$2$}  -- (c) node[above right,scale=0.8,xshift=-8pt] {$3$}  -- (d) node[below right,scale=0.8] {$4$}  -- cycle;
    \draw[dashed] (b) --++ (67.3801:1);
    \draw[dashed] (d) --++ (0:2);
    \draw[dashed] (c) --++ (-21.8014:3);
    \draw[dashed] (c) --++ (-45:-1.7);
    \draw[fill=black] (0.6765,1.6235) circle (2pt) node[above,scale=0.8,yshift=3pt] {$A$};
    \draw[fill=black] (3.5,0.0) circle (2pt) node[below,scale=0.8] {$B$};
    \draw[Maroon,thick] (0.6765,1.6235) -- (3.5,0.0);
    \draw[Maroon,thick] ($(0.6765,1.6235)!-0.3!(3.5,0.0)$) -- ($(0.6765,1.6235)!1.3!(3.5,0.0)$);
    \draw[thick, ->] ($(a)!0.3!(b)$) -- ($(a)!0.6!(b)$);
    \draw[thick, ->] ($(b)!0.3!(c)$) -- ($(b)!0.6!(c)$);
    \draw[thick, ->] ($(c)!0.3!(d)$) -- ($(c)!0.6!(d)$);
    \draw[thick, ->] ($(d)!0.3!(a)$) -- ($(d)!0.6!(a)$);
    \end{tikzpicture}
\end{gathered}
\end{equation}
i.e.,  one of the triangulations via one of the subspaces of the endpoint. In the case of the cosmological polytope, both the knowledge of the subspaces and of the compatibility conditions, is made possible by the markings, allowing to classify all these representations. The structure can be understood via the simple example above,
\begin{equation}
    \omega(\mathcal{Y},\mathcal{P})
    =
    \omega(\mathcal{Y},\mathcal{P}_{1A4}) + \omega(\mathcal{Y},\mathcal{P}_{3A2})
    =
    \frac{1}{q_{12} q_{34}} \left[ \frac{1}{q_{23}} + \frac{1}{q_{41}} \right]\,.
\end{equation}
The prefactor $\frac{1}{q_{12} q_{34}}$ determines the subspace through which the triangulation is performed, and one then sums over all the compatible boundaries, as represented by the terms in the square bracket.

\section[Asymptotic structure of cosmological integrals]{Asymptotic structure of cosmological integrals\\
\normalfont{\textit{Francisco Vazão}}}

The goal of this lecture is to extract IR divergences for any cosmological integral in FRW cosmology. We will describe in detail the integrals that arise in cosmology before providing an example of how to compute them. We will be able to show how to do it for a general tree-level integral, and describe an example of a one-loop integral before concluding.

\subsection{Infrared effects in FRW}
As a simple example of how infrared effects in FRW come about, we take a massless scalar with potential $V(\phi) = \lambda \phi^4$ in four spacetime dimensions in de Sitter spacetime, i.e., in $\text{dS}_{1+3}$.
The equal-time two-point function behaves roughly like
\begin{equation}
	\left\langle\phi_{\vec{k}}(\eta) \phi_{\vec{k}^{\prime}}(\eta)\right\rangle^{\text{{\tiny 1-loop}}} \sim \frac{H^2}{k^3}\left[1+\lambda \log \Big(\frac{k}{H a(\eta)}\Big) \log \left(k H a(\eta) L^2\right)\right] \,,
\end{equation}
where $L$ is the infrared cutoff, $a(\eta) \sim \frac{1}{\eta H}$ is the warp factor in de Sitter, $k$ is the magnitude of momentum of the fields and $H$ is the Hubble constant. There are two sources of infrared divergences in this expression. The first one is called a secular logarithm: it is an effect of putting QFT in an expanding spacetime, causing an accumulation of long wavelength modes in the future, as $\eta \to 0^-$. In this limit, the second term also goes to infinity. In addition, we have an IR divergence that comes from integrating the loop momentum over low energies, just like in flat space, which will occur upon an integration over $k$.

In examples like this one where infrared divergences are present, one of three things might happen: one cannot do perturbation theory, infrared divergences could cancel between different contributions, or one can resum an infinite set of diagrams to produce a sensible answer. So, some of the questions we want to answer here are: 
\begin{itemize}
    \item[(i)] Which conditions do the IR divergences have to satisfy in order to admit a resummation? 
    \item[(ii)] What are the completely IR finite observables in a general FRW cosmology?
\end{itemize}

In Starobinsky's stochastic inflation, the Focker-Planck equation tells us that the probability distribution function reaches equilibrium, and it has been shown that the leading logarithm matches the solution given by Starobinsky \cite{Starobinsky1986,Starobinsky:1994bd}. There are still some questions, however, e.g., what happens in general FRW cosmology and how to resum subleading logarithms. The goal of this lecture is to understand these questions in a more general setting, without restricting ourselves to a specific cosmology. 

\begin{QA}
    \question{There has already been a lot of work on this in de Sitter, so how does this talk relate to prevoius work?}
    The physical picture is that we are not starting in the free vacuum. There have already been many papers on this in de Sitter, showing that stochastic inflation resums logarithms and the system has a fixed point. We need to choose the correct state in the past, which we take to be the Bunch--Davies vacuum. How do we input the state? It is a fixed point (dynamical), so there is no simple pure state to input.
\end{QA}

\subsection{Cosmological integrals}
We take the warp factor of FRW to be $a(\eta) = \frac{1}{\eta^{\gamma}}$, where $\gamma=1$ corresponds to de Sitter but here we take $\gamma$ to be generic. 
Then, an integral corresponding to a graph $\mathcal{G}$ in perturbation theory takes the generic form
\begin{equation}
    \label{eq:generic}
	\mathcal{I}_{\mathcal{G}}[\alpha,\,\beta;\,\mathcal{X}]\:=\:
    \int\limits_{\mathbb{R}_{+}^{2}}
    \prod_{s\in\mathcal{V}}
    \left[
        \frac{\d x_s}{x_s}\,x_s^{\alpha_s}
    \right]
    \int_{\Gamma}
    \prod_{e\in\mathcal{E}^{\mbox{\tiny $(L)$}}}
    \left[
        \frac{\d y_e}{y_e}\,y_e^{\beta_e}
    \right]
    \,
	\frac{\mu_{d}(y_e;\mathcal{X})^{\frac{d-L-n_s}{2}}\, \mathfrak{n}_{\delta}(z,\,\mathcal{X})}{\displaystyle
            \prod_{\mathfrak{g}\subseteq\mathcal{G}}
	    \left[q_{\mathfrak{g}}(z,\,\mathcal{X})\right]^{\tau_{\mathfrak{g}}}} \,.
\end{equation}
Here, $\mu_d$ is the loop integration measure in $d$ spatial dimensions, which is polynomial in the loop integration variables $y_e$ and also depends on the external kinematics $\mathcal{X}$. The parameters $\alpha_s$ depend on cosmology and coupling as follows:
\be
\alpha = \gamma \Big[ 2 - \mfrac{(\kappa - 1)(d-1)}{2} \Big] + \sum_s \ell_s \,,
\ee
where $s$ are the sites of $\mathcal{G}$ and $\mathcal{V}$ is the set of all sites in $\mathcal{G}$.
The $\beta_e$ depend on which wavefunction we are computing.
$n_s$ is the total number of sites in $\mathcal{G}$, and $\mathfrak{g}$ is a subgraph of $\mathcal{G}$. The numerator $\mathfrak{n}_{\delta}$ is a polynomial of degree $\delta$. We have labeled with $z \equiv (x,y)$ the vector that contains the integration variables $x \equiv (x_s)_{s \in \mathcal{V}}$ and $y\equiv (y_e)_{e \in \mathcal{E}^{(L)}}$. Note that $\mathcal{E}^{(L)}$ is the set of edges of $\mathcal{G}$ that have at least one loop momentum running through it. $\Gamma$ is an integration contour that is given by the condition $\mu_d \geq 0$.

These types of integrals appeared in the previous subsections of this chapter, and our goal in this section is to analyze their infrared divergences.

\begin{mdexample}
The integration measure for the bubble integral is
\begin{equation}
    \mu_{\text{bub}} \sim \begin{vmatrix}
    0 & 1 & 1 & 1 \\ 
    1 & 0 & y_a^2 & y_b^2 \\
    1 & y_a^2 & 0 & p^2 \\
    1 & y_b^2 & p^2 & 0
    \end{vmatrix}
    = \text{Vol}^2
    \left(
    \begin{gathered}
    \begin{tikzpicture}[baseline={([yshift=-2ex]current bounding box.center)}]
    \coordinate (a) at (0,0);
    \coordinate (b) at (1,0);
    \coordinate (c) at (0.5,0.866025);
    \draw[thick] (a) -- (b) -- (c) -- cycle;
    \node[] at (0.5,-0.3) {$p$};
    \node[] at (1,0.6) {$y_a$};
    \node[] at (0,0.6) {$y_b$};
    \end{tikzpicture}
    \end{gathered} 
    \right) \,.
    \label{eq:intmeas_bub}
\end{equation}
\end{mdexample}

\subsection{The two-site chain}
In this section, we go through an example of how to derive the IR divergences of the two-site chain,
\begin{equation}
    \begin{gathered}
    \begin{tikzpicture}[baseline={([yshift=2ex]current bounding box.center)}]
    \coordinate (a) at (0,0);
    \coordinate (b) at (1.5,0);
    \draw[thick] (a) -- (b);
    \draw[thick,fill=black] (a) circle (2pt) node[below] {$x_1$};
    \draw[thick,fill=black] (b) circle (2pt) node[below] {$x_2$};
    \node[] at (0.75,0.35) {$y_{12}$};
    \end{tikzpicture}
    \end{gathered} 
\end{equation}
The integral is given by
\begin{equation}
    \mathcal{I}_2^{\mbox{\tiny $(0)$}}[\alpha]\: \equiv\:
    \int_{0}^{\infty}\frac{\d x_1}{x_1}\,x_1^{\alpha} 
    \int_{0}^{\infty}\frac{\d x_2}{x_2}\,x_2^{\alpha} 
    \frac{1}{q_{\mathfrak{g}}q_{\mathfrak{g}_1}q_{\mathfrak{g}_2}}\, .
    \label{eq:I_twosite}
\end{equation}
with
\begin{align}
	q_{\mathfrak{g}}=&\, x_1+x_2+X_1+X_2 \,, \\
	q_{\mathfrak{g}_2}=&\, x_1+y+X_1 \,, \\
	q_{\mathfrak{g}_2}=&\, x_2+y+X_2 \,.
\end{align}

We use the \emph{Newton Polytope} to analyze the different ways in which the denominator factors can go to zero. Recall that $X_1$ and $X_2$ are external energies, but $x_1$ and $x_2$ are integrated over. We therefore form the Newton polytope of the polynomial
\begin{equation}
    q_{\mathfrak{g}} = x_1^1 x_2^0 + x_1^0 x_2^1 + x_1^0 x_2^0 (X_1 + X_2)\,,
\end{equation}
as the polytope with vertices at the powers of each monomial that appears:
\begin{equation}
    \begin{gathered}
    \begin{tikzpicture}[baseline={([yshift=2ex]current bounding box.center)}]
    \coordinate (a) at (0,0);
    \coordinate (b) at (0,1);
    \coordinate (c) at (1,0);
    \draw[thick] (a) -- (b) -- (c) -- cycle;
    \draw[thick,fill=black] (a) circle (1pt) node[below,scale=0.8] {$(0,0)$};
    \draw[thick,fill=black] (b) circle (1pt) node[above,scale=0.8] {$(0,1)$};
    \draw[thick,fill=black] (c) circle (1pt) node[below right,scale=0.8] {$(1,0)$};
    \end{tikzpicture}
    \end{gathered} 
\end{equation}
To find the Newton polytope of a product of polynomials, such as the product $q_{\mathfrak{g}} q_{\mathfrak{g}_1} q_{\mathfrak{g}_2}$ appearing in the denominator of~\eqref{eq:I_twosite}, we can either multiply out all the factors or simply take the Minkowski sum of the polytopes that we found for each polynomial. That is, we sum their vertices. For the bubble example, we get,
\begin{equation}
    N(q_{\mathfrak{g}} q_{\mathfrak{g}_1}) = N (q_{\mathfrak{g}}) \oplus N (q_{\mathfrak{g}_1}) = 
    \begin{gathered}
    \begin{tikzpicture}[baseline={([yshift=0ex]current bounding box.center)}]
    \coordinate (a) at (0,0);
    \coordinate (b) at (1,0);
    \coordinate (c) at (2,0);
    \coordinate (d) at (1,1);
    \coordinate (e) at (0,1);
    \draw[thick] (a) -- (c) -- (d) -- (e) -- cycle;
    \draw[thick,fill=black] (a) circle (1pt)  (b) circle (1pt)
    (c) circle (1pt) (d) circle (1pt) (e) circle (1pt);
    \end{tikzpicture}
    \end{gathered} 
\end{equation}
and
\begin{equation}
    N(q_{\mathfrak{g}} q_{\mathfrak{g}_1} q_{\mathfrak{g}_2}) = N (q_{\mathfrak{g}}) \oplus N (q_{\mathfrak{g}_1}) \oplus N (q_{\mathfrak{g}_2}) = 
    \begin{gathered}
    \begin{tikzpicture}[baseline={([yshift=0ex]current bounding box.center)}]
    \coordinate (a) at (0,0);
    \coordinate (b) at (1,0);
    \coordinate (c) at (2,0);
    \coordinate (d) at (2,1);
    \coordinate (e) at (1,2);
    \coordinate (f) at (0,2);
    \draw[thick] (a) -- (b) -- (c) -- (d) -- (e) -- (f) -- cycle;
    \draw[thick,fill=black] (a) circle (1pt)
    (c) circle (1pt) (d) circle (1pt) (e) circle (1pt) (f) circle (1pt);
    \end{tikzpicture}
    \end{gathered} 
\end{equation}
The integral $\mathcal{I}$ will converge when the point $(\alpha,\alpha)$ is inside the Newton polytope. 
Next, we find the vectors normal to the facets of this Newton polytope, which we label with $\omega_i$:
\begin{equation}
    \begin{gathered}
    \begin{tikzpicture}[baseline={([yshift=1ex]current bounding box.center)}]
    \coordinate (a) at (0,0);
    \coordinate (b) at (1,0);
    \coordinate (c) at (2,0);
    \coordinate (d) at (2,1);
    \coordinate (e) at (1,2);
    \coordinate (f) at (0,2);
    \coordinate (mid) at (0.5,0.5);
    \fill[blue!20] (a) -- (b) -- (c) -- (d) -- (e) -- (f) -- cycle;
    \draw[thick] (a) -- (b) -- (c) -- (d) -- (e) -- (f) -- cycle;
    \draw[thick,fill=black]
    (a) circle (1pt) node[below left,scale=0.8] {$(0,0)$}
    (c) circle (1pt) node[below right,scale=0.8] {$(2,0)$}
    (d) circle (1pt) node[right,scale=0.8] {$(2,1)$}
    (e) circle (1pt) node[above right,scale=0.8] {$(1,2)$}
    (f) circle (1pt) node[above left,scale=0.8] {$(0,2)$};
    \draw[->,thick,Maroon] (mid) --++ (0,2) node[above] {$\omega_{2}$};
    \draw[->,thick,Maroon] (mid) --++ (0,-1) node[below] {$\omega'_{2}$};
    \draw[->,thick,Maroon] (mid) --++ (-1,0) node[left] {$\omega_{1}'$};
    \draw[->,thick,Maroon] (mid) --++ (2,0) node[right] {$\omega_{1}$};
    \draw[->,thick,Maroon] (mid) --++ (1.25,1.25) node[right] {$\omega_{12}$};
    \end{tikzpicture}
    \end{gathered}
\label{eq:NP}
\end{equation}
This pentagon is a nestohedron with five different directions that could possibly be divergent, each direction given a $\omega$.

We can now use the Newton polytope to understand the divergences of the integral~\eqref{eq:I_twosite}.
We embed the Newton polytope $N$ in $\mathbb{P}^2$ by putting $Z_i = \left( \begin{smallmatrix}
    1 \\ \rho_i
\end{smallmatrix} \right)$.
We also introduce the covectors
\begin{equation}
\mathcal{W}_I^{ab}\:=\:\epsilon_{\mbox{\tiny $IJ_1 J_2$}}Z^{J_1}_{a}Z^{J_2}_{b} = \begin{pmatrix}
    \lambda^{ab} \\ \bm{\omega}^{ab}
\end{pmatrix},
\end{equation}
with $\lambda^{ab} = (\alpha, \alpha) \cdot \bm{\omega} - \text{max}_{\rho} (\bm{\rho}\cdot \bm{\omega})$. The purpose of adding $\lambda^{ab}$ to the covectors is that $\lambda \geq 0$ indicates that the integral $\mathcal{I}$ will diverge.
The covectors for the two-site example are explicitly given by,
\begin{align}
        \displaystyle
        & \mathcal{W}^{'(1)}
        \,=\,
        \left( \begin{smallmatrix}
            -\alpha \\
            -1      \\     
            0       \\
        \end{smallmatrix} \right),
        \quad
        \mathcal{W}^{' (2)}
        \,=\,
        \left( \begin{smallmatrix}
            -\alpha \\
            0       \\     
            -1      \\
        \end{smallmatrix} \right),
        \\ & \hspace{-1.5cm}
        \mathcal{W}^{(1)}
        \,=\,
        \left( \begin{smallmatrix}
            \alpha-2 \\
            1      \\     
            0       \\
        \end{smallmatrix} \right),
        \quad
        \mathcal{W}^{(2)}
        \,=\,
        \left( \begin{smallmatrix}
            \alpha-2 \\
            0       \\     
            1       \\
        \end{smallmatrix} \right),
        \quad
        \mathcal{W}^{(12)}
        \,=\,
        \left( \begin{smallmatrix}
            2\alpha-3 \\
            1       \\     
            1       \\
        \end{smallmatrix} \right) \,.
\label{eq:covecs}
\end{align}
We can therefore read off the first entries that the integral $\mathcal{I}$ will be divergent if $2 \alpha - 3 \geq 0$, $-\alpha \geq 0$ or $\alpha-2 \geq 0 $. An equal sign denotes a logarithmic divergence, while a strict inequality signals a power-law divergence.

Let us take $2 \alpha - 3 \to 0$, which corresponds to a logarithmic divergence of~\eqref{eq:I_twosite}. From the covectors in~\eqref{eq:covecs}, we see that the only divergent direction is that of $\mathcal{W}^{(12)}$, since the first entry of the other covectors is always negative for this choice. In this case, the integral receives contributions from two divergent regions, which share the covector $\mathcal{W}^{(12)}$ but differ by including $\mathcal{W}^{(1)}$ or $\mathcal{W}^{(2)}$.

Next, we use sector decomposition to compute the infrared divergences of the integral. In the first sector, labeled by $\mathcal{W}^{(12)}$ and $\mathcal{W}^{(1)}$ we change integration variables to
\begin{equation}
	x_j\equiv\zeta_{12}^{-\mathfrak{e}_j\cdot\omega_{12}}\zeta_{1}^{-\mathfrak{e}_j\cdot\omega_{1}},\, \quad \text{for } j=1,2 \,,
 \label{eq:labelchange}
\end{equation}
where $\mathfrak{e}_j$ is a unit vector in the direction of $\omega_j$.
The integral from~\eqref{eq:I_twosite} can then be rewritten as
\begin{equation}
\hspace{-0.3cm}
    \mathcal{I}_{\Delta_{\text{IR}}^{(1)}}\:=\:
    \int_0^{1}\frac{\d\zeta_1}{\zeta_1}\,
    \frac{\d\zeta_{12}}{\zeta_{12}}\,
    \frac{\zeta_1^{-\lambda^{(1)}} \zeta_{12}^{-\lambda^{(12)}}}{
        \left[
            1{+}\zeta_1{+}(X_1{+}X_2)\zeta_{12}\zeta_1
        \right]
        \left[
            1{+}(y{+}X_1)\zeta_{12}\zeta_{1}
        \right]
        \left[
            1{+}(y{+}X_2)\zeta_{12}
        \right]
    }\,,
\end{equation}
with $\lambda^{(12)} = 2 \alpha-3$ and $\lambda^{(1)} = \alpha-2$. The divergence when $\lambda^{(12)} \to 0$ comes from the integration region where $\zeta_{12} \to 0$. Thus, when computing the leading divergence of $\mathcal{I}_{\Delta_{\text{IR}}^{(1)}}$, the change of variables in~\eqref{eq:labelchange} allows us to drop subleading terms in $\zeta_{12}$. The divergent part is therefore simply given by 
\begin{equation}
        \mathcal{I}^{\text{\tiny div}}_{\Delta_{\text{IR}}^{(1)}}
        \:=\: 
        - \frac{1}{\lambda^{(12)}}\int_0^1\frac{\d\zeta_{1}}{\zeta_1}\,
        \frac{
            \zeta_1^{1/2}
        }{
        \left(1+\zeta_1\right)
        } \,,
\end{equation}
where we have evaluated the coefficient of the pole in $\lambda^{(12)}$ at $\lambda^{(12)}=0$, resulting in $\lambda^{(1)}\vert_{\lambda^{(12)}=0} = -1/2$. The remaining integral is straightforward to evaluate, and we obtain
\begin{equation}
    \begin{split}
        \mathcal{I}^{\text{\tiny div}}_{\Delta_{\text{IR}}^{(1)}}
        \:&=\: 
        - \frac{\pi}{2 \, \lambda^{(12)}} \,.
    \end{split}
\end{equation}

The divergent contribution can be rewritten by restoring projective invariance as
\begin{equation}
        \mathcal{I}^{\text{\tiny div}}_{\Delta_{\text{IR}}^{(1)}}
        \:=\: 
        - \frac{1}{2 \lambda^{(12)}}\int_{\mathbb{R}^+_2}
    \frac{\d x_1}{x_1}\,
    \frac{\d x_2}{x_2}\,
    x_1^{\alpha-1}\,
    x_2^{\alpha-1}
    \frac{1}{\mbox{Vol}\{\mbox{GL}(1)\}}
    \frac{1}{x_1+x_2} \,,
\end{equation}
with $\alpha=3/2$,
which we can interpret as an integral corresponding to a graph with a single site of weight $x_1+x_2$. 

When adding the two divergent sectors, we can therefore write the total divergent contribution as
\begin{equation}
    \mathcal{I}^{\text{\tiny div}}_{\Delta^{(1)}}+\mathcal{I}^{\text{\tiny div}}_{\Delta^{(2)}}
    \:=\: -
    \frac{1}{\lambda^{(12)}}
    \int_{\mathbb{R}^+_2}
    \frac{\d x_1}{x_1}\,
    \frac{\d x_2}{x_2}\,
    \frac{x_1^{\alpha}\,
    x_2^{\alpha}\,}{\mbox{Vol}\{\mbox{GL}(1)\}}
    \Omega(x_1,x_2,X_1=0,X_2=0)\,,
\end{equation}
with $\alpha=3/2$.

Note that we can bypass the steps above if we instead form a graph associahedron for the tree graphs. For example, the associahedron for the two-site chain is:
\begin{equation}
    \includegraphics[scale=0.25]{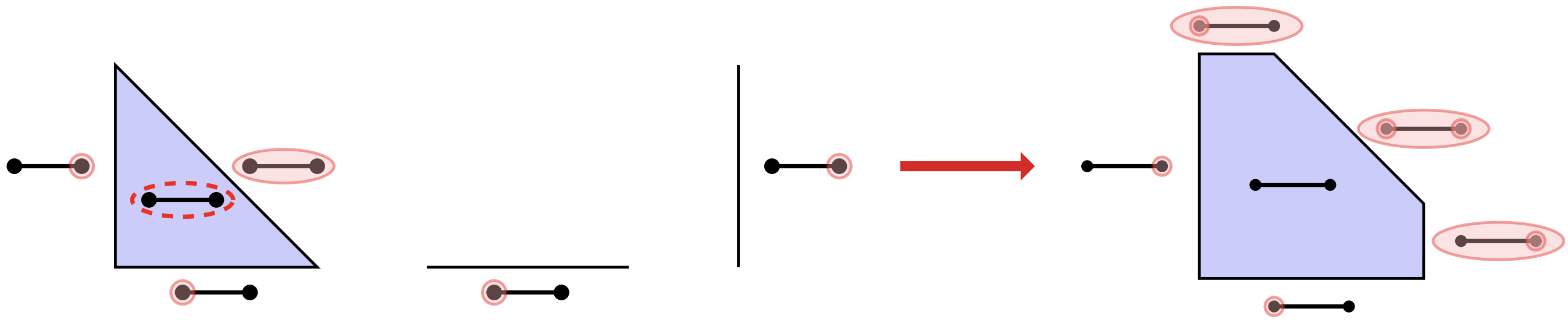}
\end{equation}
Comparing this diagram with~\eqref{eq:NP} shows that we have recovered the Newton Polytope for the graph.

\subsection{General tree graphs}
We would now like to generalize the construction in the previous section to any tree graph. A generic tree graph is given by the expression,
\begin{equation}
	\mathcal{I}_{\mathcal{G}}^{\mbox{\tiny $\text{tree}$}}[\alpha,\,\tau,\,\mathcal{X}]\:=\:
    	\int\limits_{0}^{+\infty}
    	\prod_{s\in\mathcal{V}}
    	\left[
	    \frac{\d x_s}{x_s}\,x_s^{\alpha_s}
    	\right]
    	\frac{
		\mathfrak{n}_{\delta}(x,\,\mathcal{X})
   	}{\displaystyle
      		\prod_{\mathfrak{g}\subseteq\mathcal{G}}
     		\left[
			q_{\mathfrak{g}}(x,\mathcal{X})
      		\right]^{\tau_{\mathfrak{g}}}\,
    	} \,,
\end{equation}
where we have used the same definitions as around~\eqref{eq:generic}.

Note that the Newton polytope corresponding to each denominator factor $\mathcal{N}(q_{\mathfrak{g}})$ is a simplex. Thus, we can form their Minkowski sum as
\begin{equation}
	 \mathcal{N}_{\mathcal{G}}\: \equiv\:\bigoplus_{\mathfrak{g}\subseteq\mathcal{G}}\mbox{Re}\{\tau_{\mathfrak{g}}\}\Sigma[\mathfrak{g}] \,.
\end{equation}
Schematically, it can be realized by taking the highest-dimensional simplex and truncating it with the lower-dimensional ones. Then, for all the facets that do not identify with the coordinate planes, one can associate a nested tubing consisting of all tubings which contribute in the normal direction to the facet (as illustrated in \ref{fig:3_site_Newt}). 
With the Newton polytope for the graph at our disposal, we can write down the co-vectors, which are given by
\begin{equation} 
	W^{\mbox{\tiny $(j_1\ldots j_{n_s^{\mbox{\tiny $(\mathfrak{g})$}}})$}}\:=\:
	\begin{pmatrix}
		\sum_{s}\alpha_s -(\#\text{ number of nested tubings})\\
		\mathfrak{e}_{j_1\ldots j_{n_s^{\mbox{\tiny $(\mathfrak{g})$}}}}
	\end{pmatrix}\, ,
\end{equation}
where $\mathfrak{e}_{j_1\ldots j_{n_s^{\mbox{\tiny $(\mathfrak{g})$}}}}$ is a vector with $1$ in the entries $j_1\ldots j_{n_s^{\mbox{\tiny $(\mathfrak{g})$}}}$ and zero in all other entries.
Thus, we have found a diagrammatic starting point for using Sector decomposition.

\begin{mdexample}
    The Newton polytope of the three-site chain is formed as follows:
    \begin{equation}
    \includegraphics[scale=0.20]{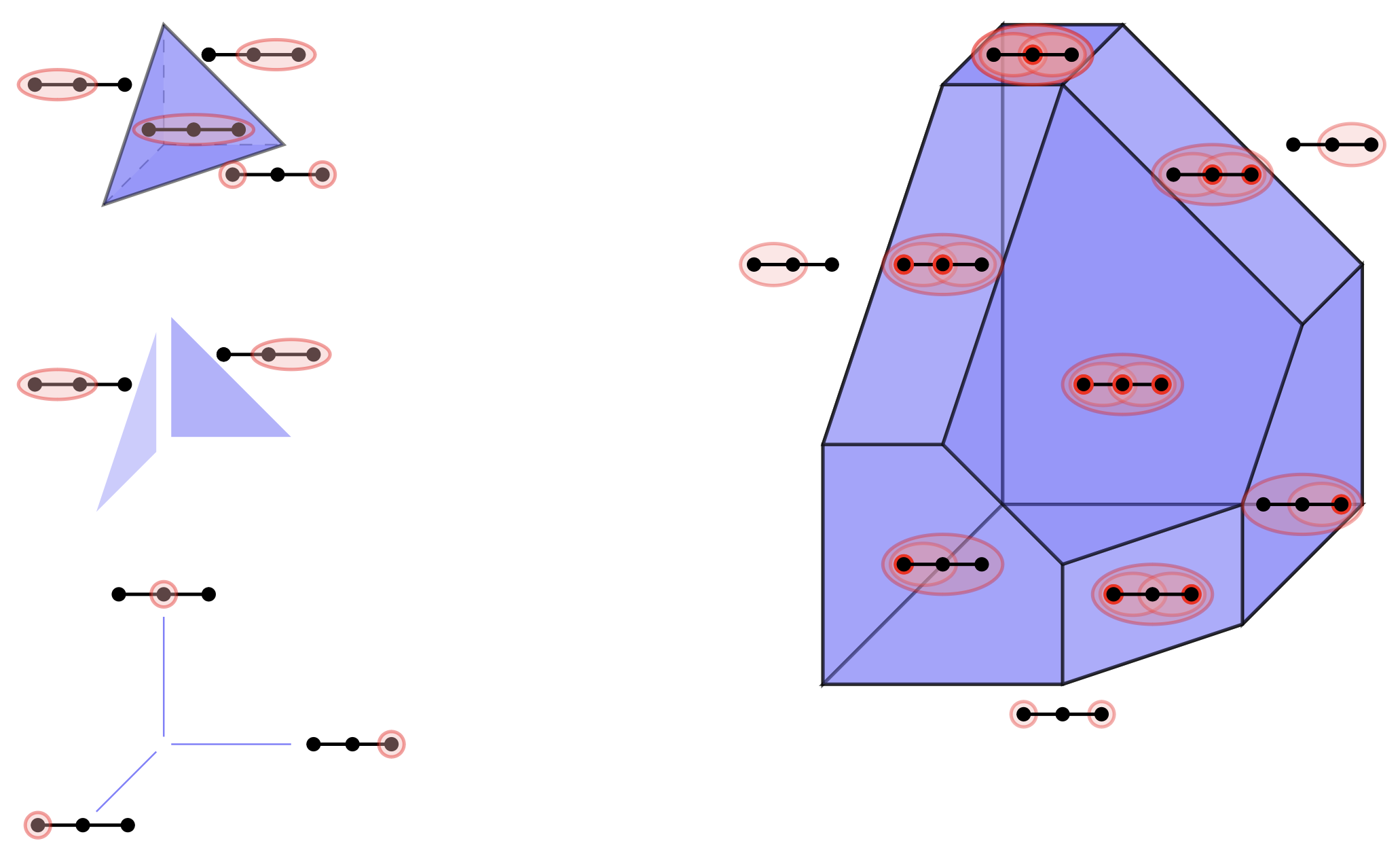}
    \label{fig:3_site_Newt}
\end{equation}
\end{mdexample}

\subsection{Loop diagrams}
In this section, we look at an example of a one loop two-site diagram, which we refer to as bubble diagram. Its expression is
    \begin{equation}
    \mathcal{I}_{\mathcal{G}}^{\mbox{\tiny $\text{bubble}$}}\:=\:
    \prod_{s\in\mathcal{V}}
    \left[
        \frac{\d x_s}{x_s}\,x_s^{\alpha}
    \right]
    \int_{\Gamma}\prod_{e\in\mathcal{E}}
    \left[
        \frac{\d y_e}{y_e}\,
        y_e^{\beta}
    \right]
    \frac{\left[-\mu(y^2,P^2)\right]^{\frac{d-3}{2}}}{P^{d-2}}\,
    \Omega\left(x,y;\mathcal{X}\right) \,,
    \end{equation}
where we use the same definitions as around~\eqref{eq:generic} and the integration measure was given in~\eqref{eq:intmeas_bub}. The denominator factors are explicitly given by
\begin{equation}
    \Omega\left(x,y;\mathcal{X}\right)\:=\:
    \frac{\mathfrak{n}_{\delta}}{
    q_{\mathcal{G}}^{\tau_{\mathcal{G}}}
    q_{\mathfrak{g}_a}^{\tau_{\mathfrak{g}_a}}
    q_{\mathfrak{g}_b}^{\tau_{\mathfrak{g}_b}}
    q_{\mathfrak{g}_1}^{\tau_{\mathfrak{g}_1}}
    q_{\mathfrak{g}_2}^{\tau_{\mathfrak{g}_1}}
    } \,,
\end{equation}
with
\begin{equation}
    \begin{split}
        &
        q_{\mathfrak{g}_a}\:= x_1{+}x_2{+}2y_a{+}X_1{+}X_2,
        \quad
        q_{\mathfrak{g}_b}\:= x_1{+}x_2{+}2y_b{+}X_1{+}X_2,\\
        &
        q_{\mathfrak{g}_1}\:= x_1{+}y_a{+}y_b{+}X_1,
        \quad
        q_{\mathfrak{g}_2}\:= x_2{+}y_a{+}y_b{+}X_2
        \quad
        q_{\mathcal{G}}\:= x_1{+}x_2{+}X_1{+}X_2 \,.
    \end{split}
\end{equation}
The integration contour is
\begin{equation}
    \Gamma = \{ y^2: \mu(y^2) \geq 0 \} \,,
\end{equation}
which from~\eqref{eq:intmeas_bub} gives the following region of integration:
\begin{equation}
    \begin{gathered}
    \begin{tikzpicture}[baseline={([yshift=1ex]current bounding box.center)},thick]
    \coordinate (a) at (0,0);
    \coordinate (b) at (0,3);
    \coordinate (c) at (3,0);
    \draw[->] (a) -- (c) node[right] {$y_a$};
    \draw[->] (a) -- (b) node[above] {$y_b$};
    \fill[black!30] (2,1) -- (1,0) -- (0,1) -- (1,2) -- (2,3) -- (3,3) -- (3,2) -- cycle;
    \draw[fill=black] (1,0) circle (0.05) node[below] {$P$};
    \draw[fill=black] (0,1) circle (0.05) node[left] {$P$};
    \end{tikzpicture}
    \end{gathered}
\end{equation}
It is clear from the integration region that whenever we set one of the edge variables to zero the other is set to $P$. This can also be seen from the perspective of the integration measure, which is the squared volume of a simplex. For the bubble we have a triangle, by setting one edge to zero the other two edges collapse into each other. For higher site graphs a similar behavior is seen, setting one edge to zero leads all others to some finite value which is a function of the external kinematics. On the other end, if we take the value of one edge variable to infinity, then all others are forced to go to infinity. This essentially fixes all the singularities of the loop integration for our bubble integral, they are $\{y_a\to 0, y_b\to P\},\, \{y_a\to P, y_b\to 0\}$ and $\{y_a\to \infty, y_b\to \infty\}$. For each of these points we can analyze their neighbor region, with a suitable change of variables, and perform sector decomposition (together with the site integration) to extract the divergences.

\newpage
\bibliographystyle{jhep}
\bibliography{references}

\end{document}